\documentclass[pre,aps,twocolumn,superscriptaddress]{revtex4-2}

\makeatletter

\usepackage{graphicx}
\usepackage{mathtools} 
\usepackage{float}

\newcommand*{\addFileDependency}[1]{
\typeout{(#1)}
\@addtofilelist{#1}
\IfFileExists{#1}{}{\typeout{No file #1.}}
}\makeatother

\begin{document}

\title{Designing 3D multicomponent self-assembling systems with signal-passing building blocks}

\author{Joshua Evans}
\affiliation{School of Molecular Sciences and Center for Molecular Design and Biomimetics, The Biodesign Institute, Arizona State University, 1001 South McAllister Avenue, Tempe, Arizona 85281, USA}

\author{Petr \v{S}ulc}
\affiliation{School of Molecular Sciences and Center for Molecular Design and Biomimetics, The Biodesign Institute, Arizona State University, 1001 South McAllister Avenue, Tempe, Arizona 85281, USA}
\affiliation{School of Natural Sciences, Department of Bioscience, Technical University Munich, 85748 Garching, Germany}

\begin{abstract}
We introduce allostery-mimetic building blocks model for self-assembly of 
3D structures. We represent the building blocks as patchy particles, where each binding site (patch) can be irreversibly activated or deactivated by binding of the particle's other controlling patches to another particle. We show that these allostery-mimetic systems can be designed to increase yields of target structures by disallowing mis-assembled states, and can further decrease the smallest number of distinct species needed to assemble a target structure. Next, we show applications to design of a programmable nanoparticle swarm for multifarious assembly: a system of particles which stores multiple possible target structures, and a particular structure is recalled by presenting an external trigger signal. Finally, we outline a possible pathway to realization of such structures at nanoscale using DNA nanotechnology devices.

\end{abstract}

\maketitle

\section{Introduction}
Self-assembly is a key process in all biological systems. In cells and bacteria, it results in structures of remarkable complexity, capable of many different functions, such as structural, enzymatic, locomotion and others. The field of bionanotechnology has long looked to biology for inspiration, and has aimed to engineer biomimetic systems that are able to self-organize into nanoscale devices and structures. 

The promising approaches of self-assembling designed systems that have been realized experimentally include DNA-coated colloids \cite{macfarlane2010establishing} and assemblies from artificially designed proteins \cite{li2023accurate}. DNA nanotechnology has emerged as one of the most promising branches of bionanotechnology \cite{seeman2017dna}. It uses single-stranded DNA strands that self-assemble into building blocks such as DNA tiles \cite{rothemund2004algorithmic} or DNA origami \cite{rothemund2006folding}. These building blocks can then be further connected by programmable interactions mediated through single-stranded overhangs to construct multiscale assemblies that can comprise repeating 2D patterns such as tilings \cite{rothemund2004algorithmic} or periodic 3D lattices \cite{tian2016lattice}. There has been a growing interest in creating finite-size assemblies \cite{tikhomirov2017fractal}, which have promising application, e.g.~as a template for precise positioning of cargo in 2D or 3D with promising applications in diagnostics, therapeutics, as well as for construction of breadboards for nanomanufacturing. Recent advances of such finite-size assemblies have shown 2D tilings, assembled from individual 2D DNA origami of up to micrometer-scale side length \cite{wintersinger2023multi}. 

Several approaches have also been used to realize finite-sized 3D assemblies. Successful examples include shape-complementary DNA origami (designed to that adjacent origami fit into each other akin to puzzle pieces) that have realized different assemblies, including capsid structures \cite{sigl2021programmable}. Similarly, self-assembled protein complexes have been designed, including engineered protein capsids \cite{bale2016accurate}. The 3D finite-size assembly rely on symmetry, and typically consists of homomeric building blocks that are designed to form self-limiting assembly. While structures that require few distinct building blocks are suitable for assembly of periodic array or lattices \cite{tian2016lattice}, finite size assemblies either rely either on self-limiting approach, such as an accumulation of strain \cite{berengut2020self} or geometric constraints \cite{sigl2021programmable} that ensure assembly into a given target structure. Alternatively, the designs have to use large number of distinct blocks to assemble the target structure. In one possible limit, each building block of a target structure is unique (so called {\it fully addressable} solution). Such approach is employed by the DNA slats assembly approach \cite{wintersinger2023multi}, and also is employed by DNA origami and DNA single-stranded tiles \cite{ke2012three}. 

\begin{figure}
    \centering
    \includegraphics[width=0.5\textwidth]{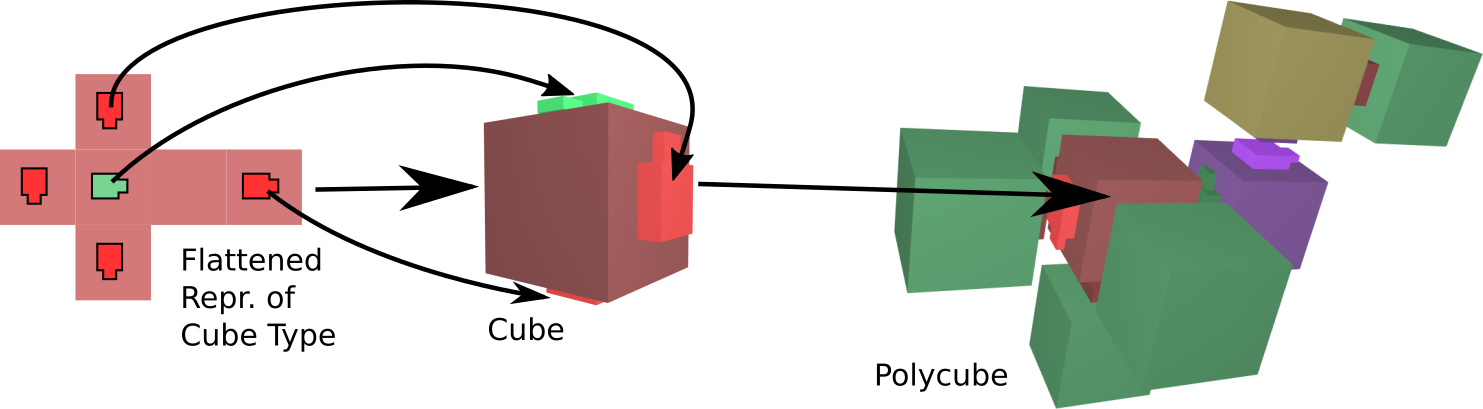}
    \caption{A schematic illustration of the polycube model. The red cube is shown unfolded on the left to fully show its patch positions, orientations, and colors. The patches are represented by the polygons on the cube faces. 
    The outset (dent) in the patch polygon indicates the patch's orientation (north, east, south, west). 
    The cube is shown in three dimensions in the center. The diagram on the right shows a polycube constructed from multiple cubes, including the red cube. Cube color indicates the cube type. Cubes of the same type have the same arrangement of patches, although the entire cube can be rotated (as the green cubes are in this example) but not mirrored. Complementary patch colors can bind to each other (e.g. the dark-red and light-red patches in this example), if they have the same orientation (e.g. cubes are rotated in such a way the dents point in the same direction). }
    \label{fig:polycube-intro}
\end{figure}

However, all these approaches rely on {\it "static"} interactions, e.g. binding sites with specificity that determine which building blocks can interact, based on whether they have complementary binding sites. These short-range binding sites are always available for binding whenever the binding partners are within a cutoff distance and at the right orientation. For proteins, the binding sites correspond to interaction domain. However, it currently remains challenging to design larger number of such orthogonal binding sites. All the bindings sites are always available, and the systems are often assembled in near-equilibrium conditions. One way to modulate such assemblies is e.g.~to introduce annealing protocol, where temperature is changed (or even cycled according to specific protocol, which can also be based on a feedback of the current state of the system) \cite{whitelam2020learning,bupathy2022temperature}. However, it is not always possible to modulate external temperature, e.g. for assemblies that happen \textit{in vivo}. Other approaches to introduce further from equilibrium condition include spatial or temporal separation of assembly (where different blocks are added at different time, or spatially separated). Recent theoretical works have further introduced the concept of non-reciprocal interactions, which can modulate non-equilibrium transitions between different assembled structures \cite{osat2023non}. However, such theoretical concepts still remain challenging to implement experimentally.  

\begin{figure}
    \centering
    \includegraphics[width=0.5\textwidth]{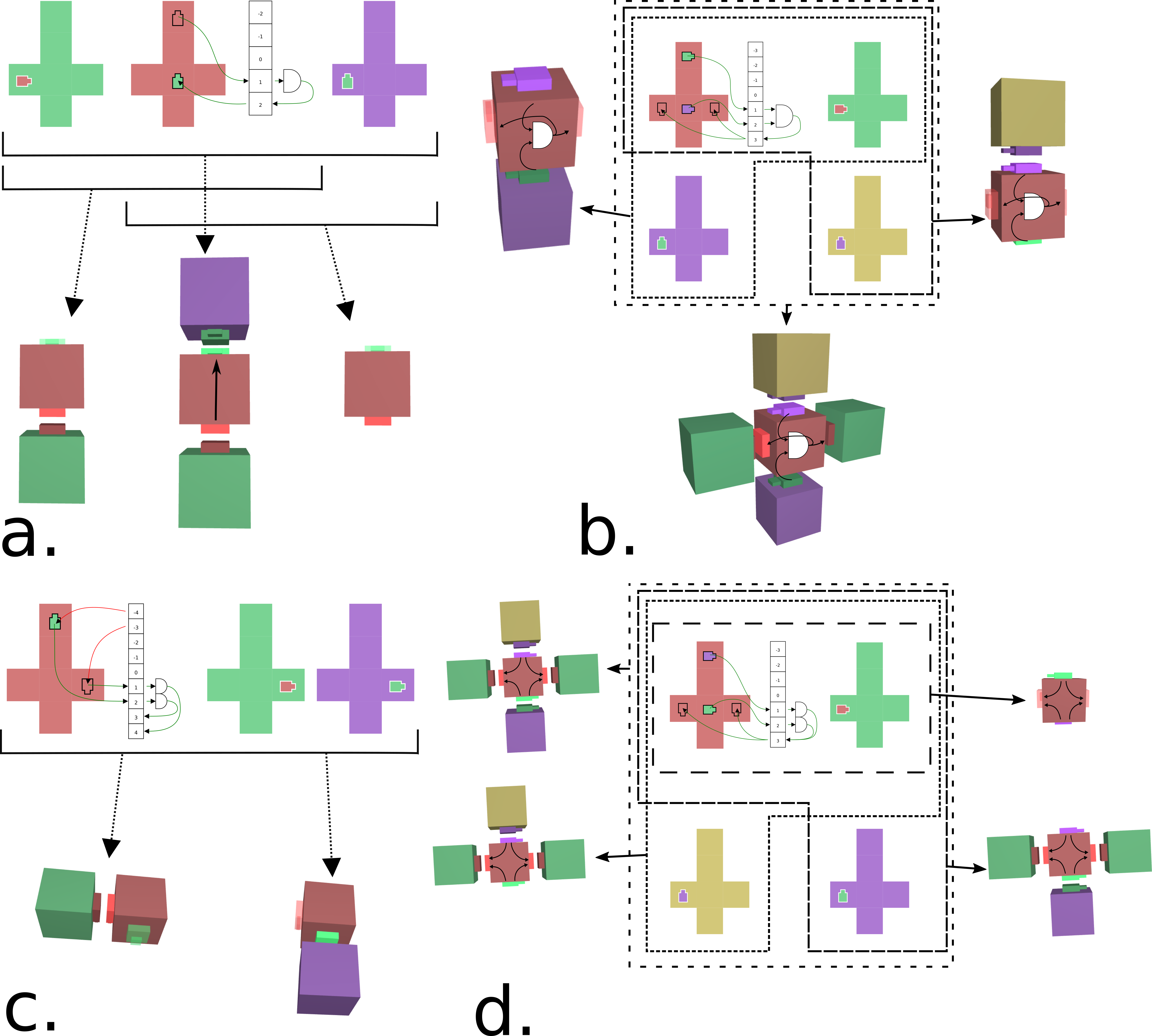}
    \caption{Schematic examples of allostery-mimetic particles:
    {\bf a)} Binding to the red patch causes the allosteric operation (the semicircle) to fire and activate the green patch. As a result of this allostery-mimetic behavior, the red cube cannot bind to the purple cube unless it first binds to the green cube. bottom-left: when the purple cube is absent, the red and green cubes are able to bind normally. If the purple cube is added (bottom-center), the particle set will form a simple linear trimer. However, if the green cube is absent (bottom-right) the green patch on the red cube will remain inactivated and no interactions will occur.
    {\bf b)} A different particle set containing a more complex logical circuit. The semicircle with two inputs is an and-gated operation, which fires only if both inputs (here, the boxes labelled 1 and 2) are True. Binding of both the yellow and purple particles activate the two red patches, allowing the two green particles to bind. Hence, the particles will form dimers (right and left) if either the yellow or purple particles are present, but will only form the cross (bottom) if both are present. {\bf c)} A demonstration of how we can construct allosteric or-gates. When the red patch on the red cube binds, it inactivates the green patch, which disables binding of the red cube to the purple cube. Binding of the green patch on the red cube functions similarly in reverse. {\bf d)} We construct an or-gate by assigning two patches to activate the same patch. Binding of the green or purple patches on the red cube sets the internal particle attribute represented by the box labelled "3" to $T$, activating the red patches. The polycubes show that if the purple, yellow, or both cube types are present, the red patches activate and allow the green cubes to bind (cross and T shapes). If neither purple nor yellow cubes are present, no activation occurs and no structure forms. (monomer in top-right).}
    \label{fig:simple-allostery-explain}
\end{figure}

There has also been an increased interest in exploring, both theoretically \cite{murugan2015multifarious,sartori2020lessons,osat2023non} and experimentally \cite{evans2022pattern} {\it multifarious} assemblies: designs where a limited set of distinct building blocks can assemble into multiple different structures. In such a case, the system is steered towards target assembly by either providing an initial seed, or biasing the system by different concentrations or interaction strengths that promote one particular target structure from multiple possible ones.

In our prior work \cite{bohlin_designing_2023}, we have explored the trade-offs between the complexity of the interaction blocks and the yield of assembled multicomponent structures. As a model system, we use polycubes (Fig.~\ref{fig:polycube-intro}): cubic-shaped building blocks with up to six interaction sites (patches). Each path is assigned a color, and only patches with compatible colors can form a bond. Each color only has one other compatible color. We quantified the complexity of a particular assembly set in terms of the total number of distinct building block species and the total number of colors (distinct orthogonal interaction interfaces). We have developed a new algorithm, called SAT-assembly \cite{russo2022sat,romano2020designing,bohlin_designing_2023}, to find solutions (in terms of assigning interaction matrix to complementary interaction sites) that have {\it minimum complexity}: the smallest number of distinct building blocks that still reliably assemble into the target shape, and avoid misassembled competing undesired structures. The SAT-assembly algorithm is able to quickly scan the design space (in terms of assigning colors to patches and designing interactions between colors). With this approach we can find sets of building blocks that can satisfy all the bonds in the desired target structure, and at the same time cannot form all bonds in undesired competing structures that have been identified in simulations (carried out either on a lattice or in a bulk using molecular dynamics simulation). The SAT-assembly method relies on mapping the inverse design problem into a Boolean Satisfiability Problem (SAT), formulated in terms of binary variables and logic clauses, for which efficient solver tools are available \cite{een2003extensible}. The method is described in detail in \cite{russo2022sat,bohlin_designing_2023}, and besides the polycube systems has also been applied to design of 3D lattices \cite{romano2020designing,rovigatti2022simple,liu2023inverse} and capsids \cite{pinto2023design}.

We here introduce a modified model, "allosteric" polycubes. These polycubes add new functionality to the model of a polycube, inspired by the signal-passing DNA tiles \cite{padilla2015signal} and allosteric modulation in proteins. In this model, interaction patches in the building blocks can be activated by binding of some other patch, akin to protein allostery where binding of a ligand in one region can change a distant region and make it available for binding of a new substrate. Besides this direct allostery-mimetic coupling between patches, we further include in our model the capability to perform a logic-gated activation: binding to several input patches then goes through AND or OR gate, which controls activation of remote output patches. The allosteric polycubes are shown in Fig.~\ref{fig:polycube-intro}. 

Our model represents a minimalistic extension of the polycube model which only had static interaction sites. Several models of "nanoscale" robotics systems with self-assembly mechanisms have been previously theoretically proposed, such as programmable diffusing robots "amoebots" \cite{derakhshandeh2014amoebot}. Such models are capable of more complex neares-neighbor communication, which allows them to achieve more complex task, but our currently available experimental capabilities do not allow for realization of such complex function at nanoscale. Simpler than most proposed programmable matter system, our allosteric polycubes model systems might still be experimentally tracktable and realizable with current existing tools in DNA or protein nanotechnology. The proposed allostery-like signal passing through nanostructure and through logic gate can be in principle realized through strand displacement cascades or mechanical rearrangements designed in nanostructures.

Prior works have studied 2D assemblies of signal-passing tiles \cite{hendricks2013signal,padilla2014asynchronous,cantu2020signal}, inspired by signal-passing DNA tiles where binding to one end unblocks the other end \cite{padilla2015signal}. Our model can be understood as extension of this model to 3D, with added computational capacity to the tiles (AND and OR gates that control activation of the remote patch) and new examples of applications, towards design and eventual application of realization of programmable nanoparticle swarms, which can be potentially realize experimentally. 

In this work, we first introduce in detail the allostery-mimetic polycubes model, and then study different target assemblies that can be achieved with this mechanism. We show that compared to a static model, the total number of different particle species that self-assemble into a target shape can be further reduced if we introduce the signal passing capacity in them. Additionally, we also show that allostery-mimetic polycubes can be used to specifically remove some misassembled states that have been observed in static particles, thus improving the overall yield.
Next, we explore application to multifarious assemblies. We study the capacity of our system to encode multiple structures at once, and its ability to recall one of the stored structures by providing a trigger signal, and show several examples of designed systems where such assembly is verified in simulation. Finally, we outline possible realization of such signal passing logic with DNA origami \cite{rothemund2005design} nanostructures.




\section{Methods}
\subsection{Design of Allostery-Mimetic Particle Sets}

\subsubsection{The Stochastic Lattice Assembler}
For design purposes, particles were abstracted as cubes within a polycube model derived from the one described in Ref.~\cite{bohlin_designing_2023}. The polycube model abstracts each particle as a cube (see Fig.~\ref{fig:polycube-intro}), which can occupy an integer position in a 3D lattice. Each cube can have up to six patches, each of which is positioned on one of the six faces. Each patch has a "color". Each color is assigned one other complementary color. In the torsional version of the polycubes model, each patch also has an orientation (North, East, South, West), which represents one of the possible four orientations that the patch can have. Two patches can bind only if they are positioned in identical orientation and their colors are complementary. Cubes can be rotated at multiples of $\frac{\pi}{2}$-radian angles, but cannot be flipped mirror-wise. When the cubes are rotated, the patch orientations are correspondingly rotated. The polycube model can be simulated on a 3D lattice, e.g. with a stochastic growth algorithm where cubes can be attached from available patches of other cubes. In this model, patch binding is irreversible. The model acts as if it has an infinite reservoir of all particle types, so all valid particle types are equally likely to be attached to any given patch.

\begin{figure}
    \centering
    \includegraphics[width=0.5\textwidth]{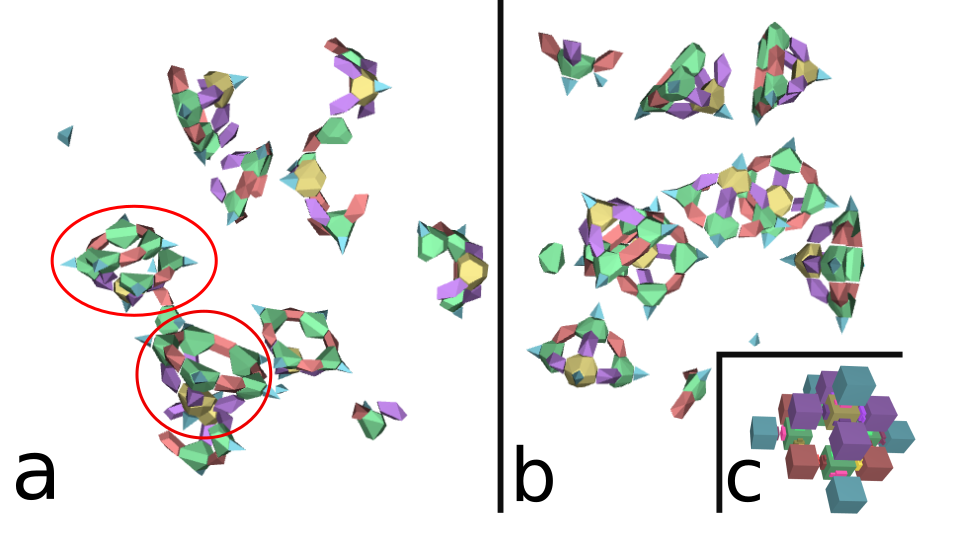}
    \caption{How allostery-mimetic signalling can improve yield of pyramid design by preventing spiral defects. The particles are stylized here as polyhedra for easier visualization; In lattice simulations they are treated as cubes, and in molecular dynamics simulation as spheres. {\bf a)} Result of a simulation of the non-allosteric pyramid. Spiral defects, where two different edges overgrow instead of connecting into a cycle of desired length, are circled in red. {\bf b)} Result of a simulation of a pyramid with allosteric design Version 1 (see Fig.~\ref{fig:pyramid-yield}). {\bf c)} The pyramid in polycubes, for reference. Colors of cubes correspond to different particle types, and correspond to colors of particles in a) and b). The stochastic assembler on the 3D cubic lattice is unable to identify spiral defects, as they are not possible on a lattice.}
    \label{fig:pyramid-yield-eg}
\end{figure}

The stochastic assembler adds particles step by step. In each step, it randomly selects a patch on a particle on the assembly that doesn't have a neighboring particle and adds a particle adjacent to it, connected by a patch bond. Patch bonding in the stochastic assembler is irreversible. The stochastic assembler continues adding particles until it either runs out of patches with unoccupied adjacent space or arrives at a predetermined step count. 

\begin{figure*}
    \centering
    \includegraphics[width=\textwidth]{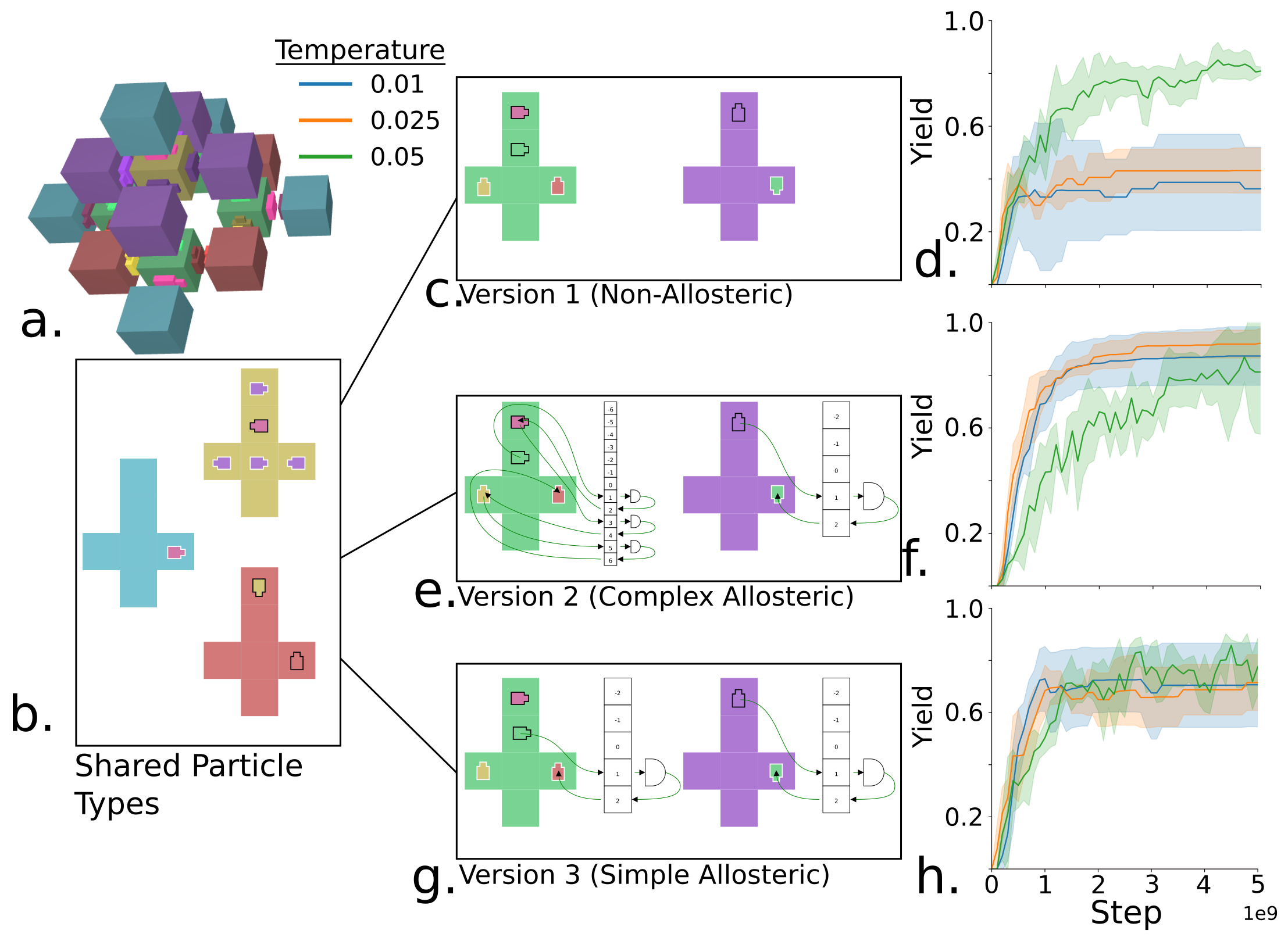}
    \caption{The pyramid design improves yield at low temperatures. {\bf a)} The pyramid structure visualized as a polycube. Cube colors correspond to different particle types, shown in {\bf b}, {\bf c}, {\bf e}, \& {\bf g}. {\bf b)} Cube types shared between the non-allosteric pyramid and the two versions of the allosteric pyramid. {\bf c)} Two cube types used in the static (non-allosteric) pyramid. {\bf d)} Yield curve for the particle set shown in {\bf c}. {\bf e)} An allosteric pyramid particle set with a complex allosteric mechanism. {\bf f)} The yield curve for the Version 2 pyramid set shown in {\bf e}. {\bf f)} A pyramid design with a simpler allosteric mechanism. {\bf g)} The yield curve for the particle set in {\bf f}}
    \label{fig:pyramid-yield}
\end{figure*}

\subsection{Signal-passing Allosteric Polycubes Model}
We extended the polycubes model with allostery-mimetic behavior. 
We consider that each particle's patches belong to either "static" or "allosteric" category. The static patches are always available for binding (active), while allosteric patches are either set in T (true/active) or F (false/deactivated) state. If they are set to T state, they are available for binding and can form bonds, whereas in F state they are inert and cannot form any bonds. In our model, a patch activation can be irreversibly switched either from F to T or from T to F. Each allosteric patch is controlled by a set of "control patches". Once the control patches form a bond with other particle, they propagate their activation signal. If an allosteric patch is controlled just by one patch, a particle binding to the control patch leads to a switch of the controlled patch state from F to T (or F to T in case of the control patch disabling the interaction), as schematically shown in Fig.~\ref{fig:simple-allostery-explain}a). If an allosteric patch is controlled by two or more control patches, their binding state (1 if patch is occupied or 0 if it is not part of a bond) is processed through a logical circuit, which then can set the allosteric patch's state from F to T (or T to F) if the output of the gate becomes 1 (see Fig.~\ref{fig:simple-allostery-explain}b).  We further elaborate on the internal workings of this model in the supplemental material.


In order to correctly assemble structures from allosteric particle sets, we introduced piecewise assembly into the stochastic assembler. In piecewise assembly, instead of individual particles the stochastic assembler adds small pieces to grow the structure. The pieces are constructed by executing the stochastic assembler for three steps before the main assembly starts. Single-particle pieces are allowed.
The rationale behind piecewise assembly is due to the fact that the for the allosteric model, the assembly order matters (it does not have associative property), and is fully explained in the supplementary material.

The code for the Polycubes stochastic assembler can be found at \texttt{github.com/sulcgroup/polycubes-clone}.

\subsection{Design Methodology}
The goal of our design approach is to find the smallest set of distinct particle species that reliably assemble into a target shape (or multiple different shapes in the case of multifarious designs). All of our non-allosteric particle sets were designed using the SAT-assembly method described in Ref.~\cite{bohlin_designing_2023}. Briefly, for a given number of distinct species and patch colors, the algorithm finds assignment of patch colors on each particle species such than when run using the stochastic assembler, the system only assembles the desired structure 100 times out of 100. Finding the smallest set in terms of number of distinct particle species and patch colors than constitutes what we call a {\it minimal set}. 

All of our designs which include allosteric patches have been designed rationally, based on our insight from observed misassembled states from the static model designs in molecular dynamics simulations, or based on our intuition when designing for particular desired behavior.

\begin{figure*}
    \centering
    \includegraphics[width=\textwidth]{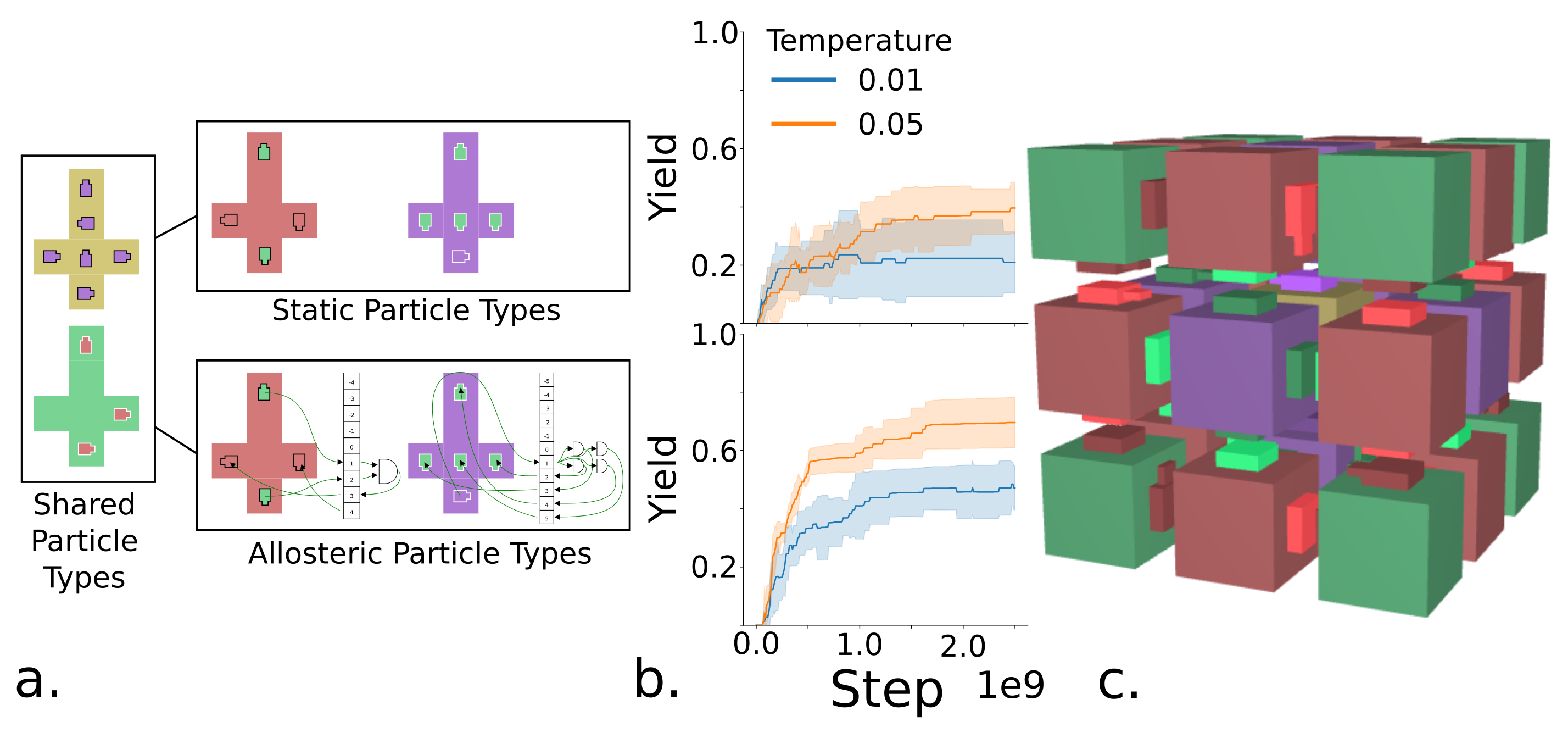}
    \caption{Compared yields of the static 3x3x3 solid cube particle set and an allosteric version of the particle set. {\bf a)} Schematics of the particle sets. The yellow and green particle types (left box) do not change from the static and allosteric sets. The particles altered between the two sets are shown in the upper and lower right boxes. {\bf b)} Yield curves for the static (top) and allosteric (bottom) particle sets. {\bf c)} A visual representation of the 3x3x3 solid cube from the Polycubes model.}
\end{figure*}

\subsection{Patchy Particle Simulation}
For the polycube designs in this work, we first verify them using a stochastic assembly algorithm on a 3D lattice, which can quickly identify possibly misassembled 
states and redesign the interactions between particles using SAT-assembly (for static interactions) or rational interaction redesign for the particles with allosteric patches.

After the assembly is verified on a 3D lattice, we use patchy particle representation and run molecular dynamics simulation of assembly of target structure. The patchy particles can e.g.~represent DNA origami nanostructures or designed proteins, and despite high level of abstraction still provide a more accurate representation than the cubes on a 3D lattice, and can be used to e.g. identify misassembled states that could hinder the assembly of the target structure \cite{liu2023inverse}.

\subsubsection{The Patchy Particle Model}
\label{sec:patchy-methods}
Patchy particle simulations were performed using an implementation of a patchy particle code from Ref.~\cite{bohlin_designing_2023}, which we further extended to include the (optional) allosteric modulation of patches. 

In brief, particles are modeled as soft spheres (with radius $0.5$ in simulation distance units) with an excluded volume interaction modeled by a repulsive Lennard-Jones potential \cite{bohlin_designing_2023}.
In order to mimic cubic particles, interaction sites (patches) were placed at orthogonal points on the surface of the spheres, at distance $0.5$ units from the particle center. In addition to position, each patch was defined by a color, an $a_1$ vector, and an $a_3$ vector. Two patches were only able to interact if the patch colors are compatible (and both patches are activated in case of allosteric particles), and the $a_1$ and $a_3$ vectors are aligned properly. 
The equation for the interaction energy between two patches is described by Eq.~\ref{eqn:patch_interaction_allo}. The interaction is similar to the one used previously in Ref.~\cite{bohlin_designing_2023}, but with the addition of the $\alpha_i$ and $\alpha_j$ terms which implement allostery. 
The patch compatibility coefficient $\delta_{ij} \in \{1,0\}$ is a value which is 1 if the two patches' color are compatible and is 0 otherwise. $\alpha_i$ and $\alpha_j$ $\in \
\{1,0\}$ are the allosteric patch activation coefficients, which are 1 if the corresponding patch is activated (in "T" state) and zero otherwise. For non-allosteric static designs $\alpha_i$ is always 1. The function $V_{pdist}$ is applied to the distance between patches $r_{ij}$. The values $\theta_a$, $\theta_b$, and $\theta_t$ are angles derived from $\Omega_{ij}$ which express the relative orientation of the two patches.

The angles $\theta_a$, $\theta_b$, and $\theta_t$ are defined in Eq.~\ref{eqn:patchy-theta}:
\begin{eqnarray}
    \label{eqn:patchy-theta}    
        cos\, \theta_a &=& \hat{r}_{ij} \cdot \hat{p_a}  \\\notag
        cos\, \theta_b &=& -\hat{r}_{ij} \cdot \hat{p_b} \\ \notag
        cos\, \theta_t &=& \bf{o}_a \cdot \bf{o}_b, 
\end{eqnarray}
where $\hat{r}_{ij}$ is the normalized vector between the centers of mass of particles $i$ and $j$ and $\hat{p_a}$ and $\hat{p_b}$ are the normalized vectors from the centers of mass of particles $i$ and $j$ to patches $a$ and $b$ respectively. Thus $\theta_a$ is the angle between the vector from particle $i$ to particle $j$ and patch $a$ (which is on particle $i$), and $\theta_b$ is simply the same but for particle $j$ and patch $b$. $\theta_t$ is the angle between the two patch orientations, which is zero if the patches are perfectly aligned.

\begin{figure*}
    \centering
    \includegraphics[width=\textwidth]{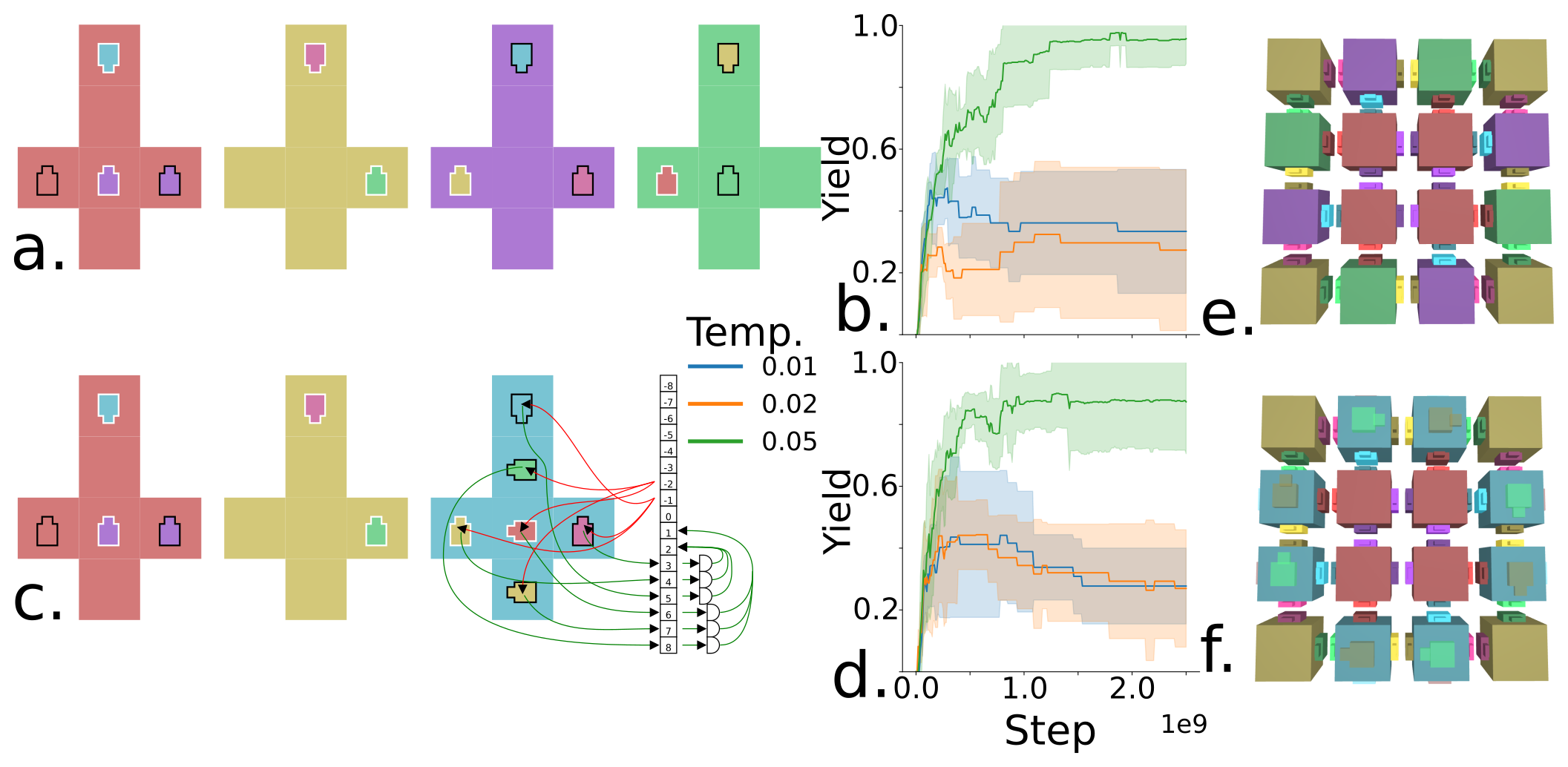}
    \caption{Assembly of 4x4 tiles with and without allostery. The yield curves show that both particle sets self-assemble at comparable rates to produce comparable yields as the design with static patches. {\bf a)} The static 4x4 tile particle set. {\bf b)} A yield curve showing assembly of the static 4x4 tile. {\bf c)} The reduced particle set, which uses allostery to lower the number of particle types required by one. It was designed by the algorithm described in \ref{fig:reducer-explain} and the accompanying text. {\bf d)} A yield curve showing assembly of the reduced 4x4 tile particle set. {\bf e)} A polycube visualization of the static 4x4 tile. Cube colors correspond to particle types in {\bf a} and {\bf b}. {\bf f)} A polycube visualization of the reduced 4x4 tile. Cube colors correspond to particle types in {\bf c} and {\bf d}.}
    \label{fig:4x4tiles}
\end{figure*}
 
\begin{align}
    \label{eqn:patch_interaction_allo}
    V_{patch}(r_{ij},\Omega_{ij},\alpha_i, \alpha_j) &= \begin{multlined}[t]
        \delta_{ij} \alpha_i \alpha_j V_{pdist}(r_p) \\ 
        \times V_{ang}(\theta_a) V_{ang}(\theta_b) V_{ang}(\theta_t)
     \end{multlined} \notag \\
\end{align}
The function $V_{\rm ang}(\theta)$ is described in Eq.~\ref{eqn:angmod}. It implements the torsional modulation that ensures  The coefficients $a$, $b$, $\Delta$, and $\Delta_c$ define the tolerance for angles between patches. In our simulations, $a = 3$. $b = 0.730604$, $\Delta = 0.2555$, $\Delta_c = 1.304631$. 
\begin{equation}
    \label{eqn:angmod}
    V_{\rm ang} (\theta) = 
    \begin{cases} 
        1 - a\theta^2 & \text{if } - \Delta < \theta \\
        b(- \Delta_c - \theta)^2 & \text{if } - \Delta_c < \theta \\
        b(\Delta_c - \theta)^2 & \text{if } \Delta_c < \theta \\
        0 & \text{otherwise}
    \end{cases}
\end{equation}

The molecular dynamics simulations are ran at different temperatures (specified in units of $\epsilon/k_{\rm B}$, where $\epsilon$ is the simulation energy unit). Particle energies were kept consistent with the simulation temperature using an Anderson-like Thermostat \cite{russo_reversible_2009, bohlin_designing_2023}.

During runtime, each time a particle formed a patchy bond, the program looks up the transition in the transition map that stores the allosteric activating rules and flipped allosteric patches's state from F to T if the activation rules are satisfied (or T to F in case of deactivating allosteric switch). The switch from F to T (or F to T) is irreversible, and even if the particle that caused the switch unbinds during the simulation, the controlled patch state can no longer switch back to its starting state. 

The simulation code was implemented with the oxDNA simulation package tool \cite{poppleton2023oxdna} and is freely available at \texttt{github.com/sulcgroup/alloassembly}.

\subsection{Evaluation of designs}
In order to determine if a design was viable before testing it on the Patchy Particle model, we used the polycube model. In Polycubes, a design was considered viable if and only if it was bounded (the stochastic assembler added a finite number of particles, then stopped) and deterministic (the stochastic assembler consistently produced the same structure, or structures for multifarious assembly).

We compute yields from patchy particle simulations using a graph matcher algorithm from the IGraph \cite{igraph} Python package. Particle cluster topology data was recorded during simulation execution and converted to structure graphs. At each timepoint, the algorithm looped through each cluster and attempted to find a subgraph of the target structure that was homomorphic to the cluster.  In the case of the multifarious designs, the algorithm was run with different target structures to identify yields of different targets. The individual cluster yield was the size fraction - the ratio of the number of particles in the cluster to the target structure size. The program then applied a filter that treated the yield of all clusters with size fractions below a preset cutoff point (0.75 in all main text figures; alternative cutoff points are explored in the supplemental). The absolute yield was computed by summing the yields of all clusters that made it through the filter. The relative yield - $0.0$ to $1.0$ values used in all figures here - was computed by dividing the absolute yield by the yield that would be produced by perfect assembly efficiency. In most of our simulations, this was 8 fully assembled structures (Exceptions include the solid cubes and stonehenge designs, as discussed later). For the wereflamingo design, we used a modified version of the subgraph isomorphism algorithm which required that clusters contain the correct particle types. The rest of the algorithm for yield computations remained the same.

\subsection{Molecular Dynamic Simulations Conditions}
With the exception of the stonehenge design simulations, molecular dynamic simulations were run in cubic simulation boxes with periodic boundary conditions. Box sizes were set so that particles were at a density of 0.1 particle types per cubic simulation unit. Particles were placed randomly in the simulation box prior to the simulation space.

The conditions of each of our molecular dynamic simulations are fully described in the supplementary materials.

\begin{figure}
    \centering
    \includegraphics[width=0.5\textwidth]{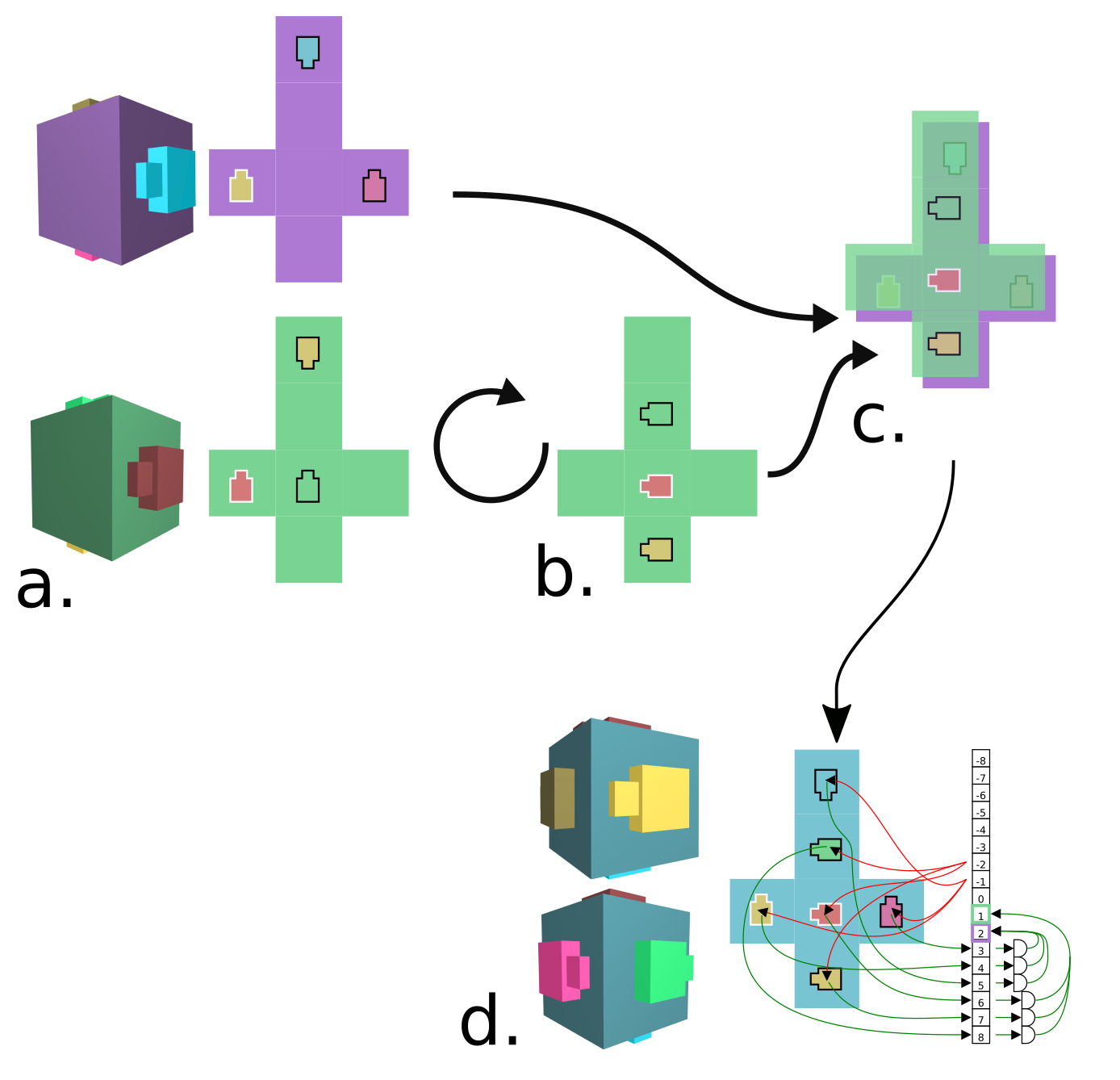}
    \caption{The cube type merge algorithm. {\bf a)} A pair of cube types to be merged. {\bf b)} The green cube is rotated so that none of its patches overlap with patches on the purple cube. {\bf c)} The cubes are overlaid. {\bf d)} The result of the merge, which has allosteric controls that ensure it acts like either the green cube or the purple cube, depending on its patch interactions. A more thorough explanation of the steps between {\bf c} and {\bf d} is provided in Suppl.~Fig.~\ref{fig:reducer-suppl}.}
    \label{fig:reducer-explain}
\end{figure}

\begin{figure}
    \centering
    \includegraphics[width=0.5\textwidth]{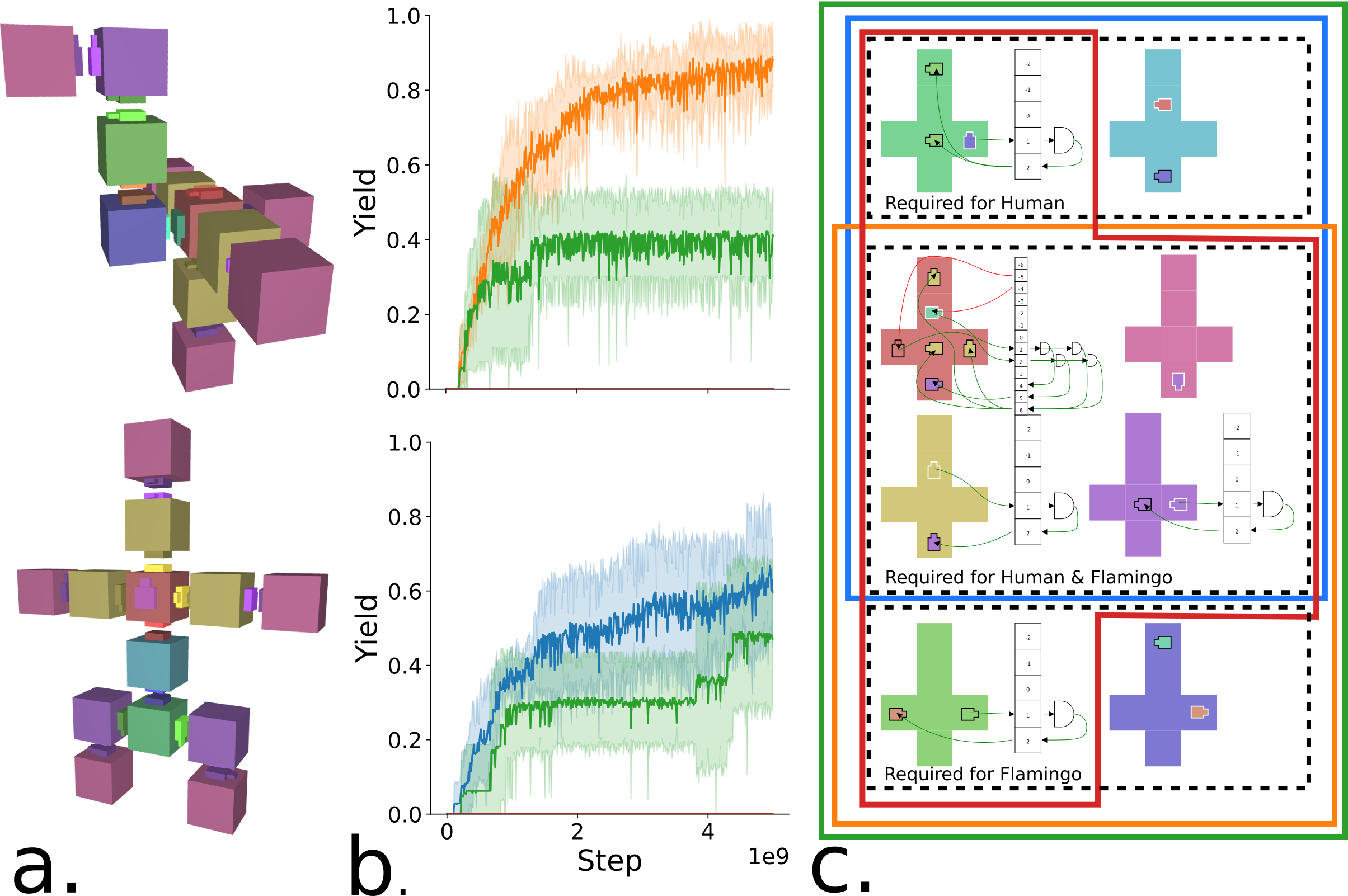}
    \caption{{\bf a)} The human and flamingo structures as polycubes. {\bf b)} Yields for formation of the flamingo structure (top row) and the human structure (bottom row). The color of the lines in the yield curves correspond to the sets of particles included in the yield curves. The green curve is the mixture, the blue curve is the human-only particle set, the orange curve is the flamingo-only particle set, and the red curve (barely visible) is the particle set that contains no trigger particles. The upper and lower plots are measurements of formation of the flamingo and human respectively, and are derived from the same simulation data. The yields were calculated taking into account particle type in order to avoid problems caused by topological similarities between the structures. {\bf c)} The particle sets that compromise the experimental groups. The colored boxes correspond to the yield curves.}
    \label{fig:wereflamingo-yields}
\end{figure}

\begin{figure}
    \centering
    \includegraphics[width=0.5\textwidth]{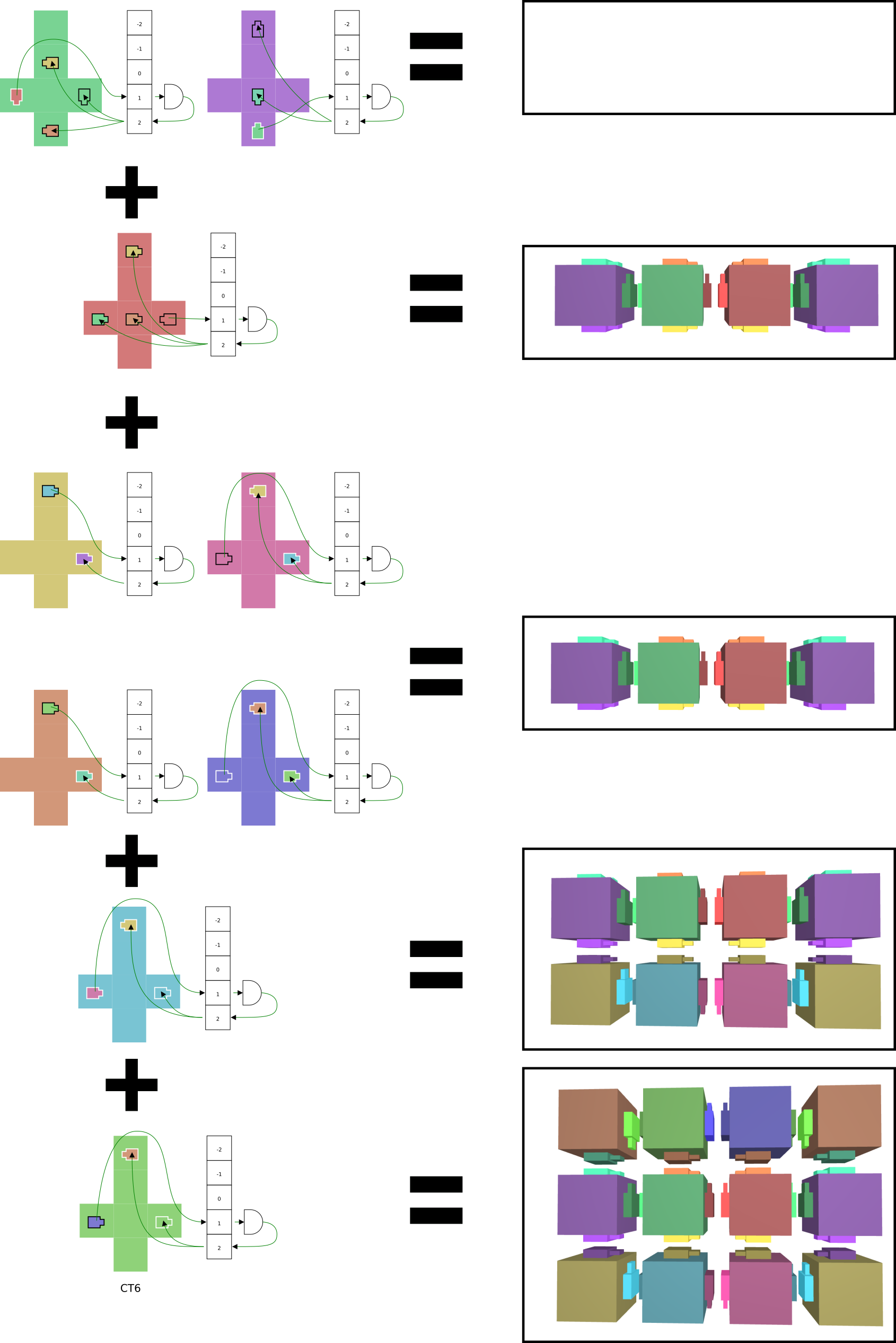}
    \caption{The triple bridge allosteric particle set. When only the green and purple particle types are present, no formation occurs. When the red particle types is added, the particles self-assemble into the "single bridge" tetramer. No additional structures are formed when the yellow, magenta, orange, and blue particle types are added, but once the cyan particle type is added the particles will continue assembling to form the double bridge.}
    \label{fig:bridge-constitutive}
\end{figure}

\section{Results and Discussion}
In the following section, we investigate different applications of the allostery-mimetic self-assembling systems. In particular, we look at i) using allosteric patches to improve yields of structures that contain cycles, ii) decreasing the number of distinct particle types needed to assemble target shape, and iii) designing "multifarious" systems, where a group of particles are directed by an external signal (addition of another particle or positioning of a seed particle) to self-assemble into one of several possible structures.
Most of the designs presented here have been designed rationally, using the stochastic assembler and molecular dynamics simulations, as described in Methods.

\subsection{Allosteric Particles Improve Yield of Low-Temperature Assemblies}

To study the effects of allostery on patchy particle assembly yield we designed allosteric particle sets for pyramid and solid cube designs.
Yield-improving allosteric designs improve yield by preventing spiral defects, as shown in Fig.~\ref{fig:pyramid-yield-eg}, where two different lines in a cycle of the self-assembling structure overgrow before they can complete the cycle. To prevent this defect, we design allosteric modulation to force the structure to grow from an origin point along a specific design path. Ideally, each particle in the design path would activate the next particle, but in practice this is neither practical nor desirable, as more allosteric controls necessarily lower the rate at which particles form structures. In practice, we used our experience to design the allosteric assembly pathways to improve yield.

We developed two different particles set designs of the pyramid particle set. The one we refer to as Version 1 has no allosteric behavior (top row of Fig.~\ref{fig:pyramid-yield}) and is used as a reference for comparison. The one we refer to as Version 2 has a complex allosteric behavior that requires a very specific ordering of particle binding. It doesn't significantly alter yield at $T=0.05$ and improves yield at $T=0.01$ and $T=0.025$ from about 40\% in Version 1 to about 90\% in Version 2. (see the middle row of Fig.~\ref{fig:pyramid-yield}). Version 3 is simplified to the core essentials, and still shows a marked increase to roughly 75\% in $T=0.01$ and $T=0.025$ yield.

We also used allosteric behavior to improve yield of the 3x3x3 solid cube described in \cite{bohlin_designing_2023}. We designed and tested four allosteric solid cube rules, of which only one demonstrated an improvement in yield. (Data for the other allosteric versions are shown in Supp.~Fig.~\ref{fig:solid-cubes-suppl}). The allosteric control forces the cube to form starting from the yellow particle type at the center (see Suppl.~Fig.~\ref{fig:solid-cube-assembly}). This helps prevent spiral defects (similar to the ones shown in Fig.~\ref{fig:pyramid-yield-eg}). 

Further evidence for allosteric behavior improving yield by preventing the formation of spiral defects is shown by adjusting the torque potential. (Eqs.~\ref{eqn:patchy-theta} \& \ref{eqn:angmod}, torque potential adjustments are fully described in the supplementary materials). We found that the difference between the yield improvement of the allosteric design over the static one improves as we widen the torque modulation, as wider torque modulation leaves more room for spiral defects. (See supp. info, Suppl. Figs.~\ref{fig:pyramid-wt-comprehensive}, \ref{fig:pyramid-x1-comprehensive}, \& \ref{fig:pyramid-x4-comprehensive}). For narrower modulation, the spiral defects are less likely to appear, and the yield of ther static design improves.

\begin{figure}
    \centering
    \includegraphics[width=0.5\textwidth]{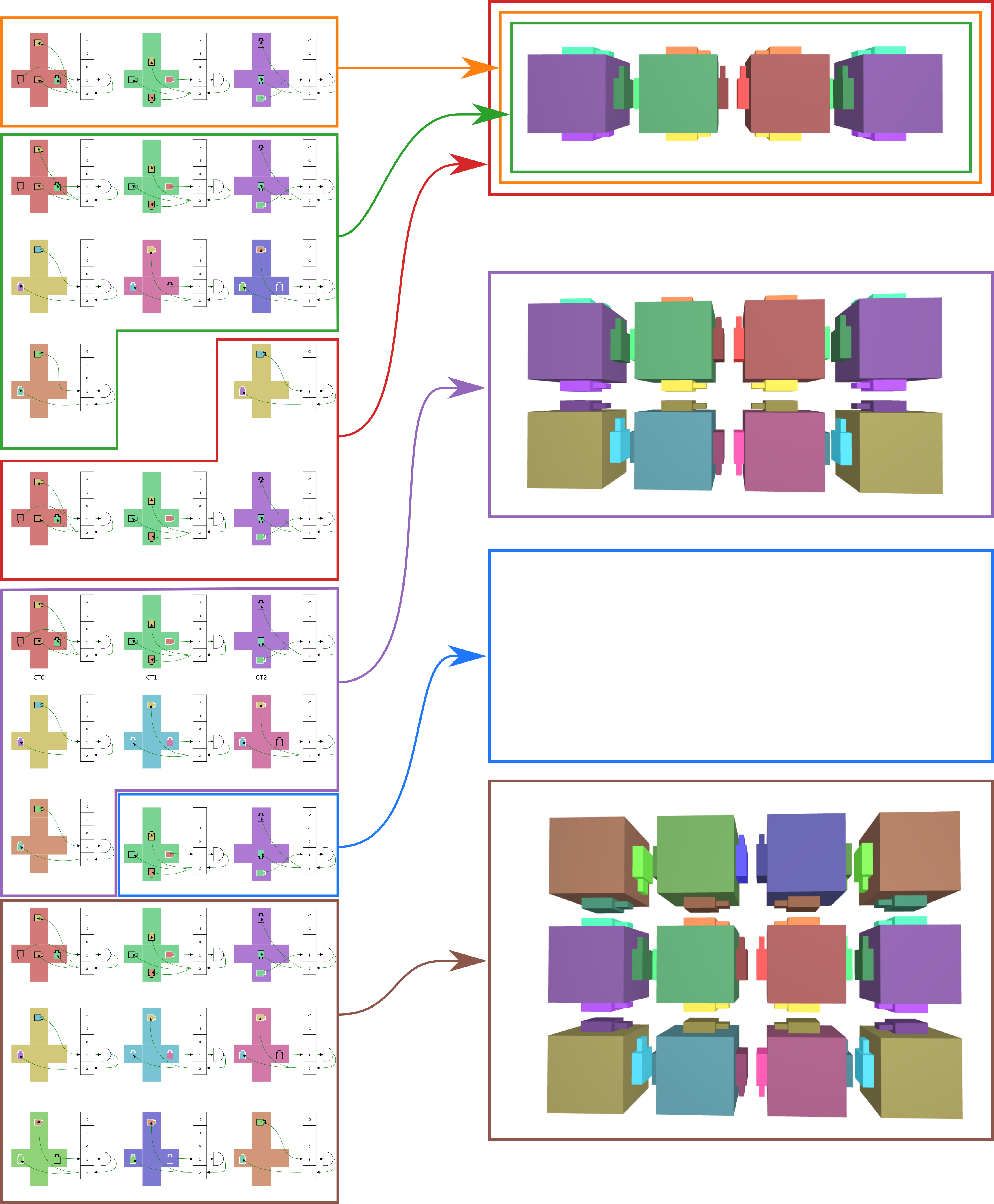}
    \caption{Particle sets and test groups for the triple bridge. Each colored box on the left shows an experimental setup containing different particle types. The structures formed in the polycubes model are shown on the right. The colors of the boxes showing the experimental setups correspond to the color coding used in Fig.~\ref{fig:bridge-yields}. Of note, the blue-outlined particles have complementary colors but will not be able to bind without other particles present due to the allosteric control over the patches. As a consequence, they have no yield.}
    \label{fig:bridge-particle-sets}
\end{figure}

In our observations, design of allosteric controls to improve yield requires that the structure have i) cycles, where allostery is used to avoid them, and ii) it has a "root" particle type which occurs exactly once in the target design, from which the assembly can then grow. We rationally designed and tested all possible allosteric wireframe cube particle sets for two particle species (Suppl.~Fig.~\ref{fig:wireframe-cubes}). The wireframe cube structure contains cycles but does not contain a central particle type that occurs only once in the assembled structure. Unsurprisingly, none of our allosteric wireframe cubes demonstrated a meaningful yield improvement over the static particle set.

Many assemblies in biology happen at constant temperature, and hence it is of interest to improve self-assembly yield for environments where one does not have a constant temperature control, such e.g. for artificial cell design or synthetic biology applications.

\subsection{Reductions in Particle Type Count using Allostery}
In order to minimize the number of different allosteric particle types that are needed for assembly of a target shape, we developed an automated process to reduce particle type counts by introducing allosteric behavior. The program loops through pairs of particle types and attempts to merge each pair. 
The cube merge algorithm is shown in Fig.~\ref{fig:reducer-explain}. In brief, the algorithm first superimposes the two cube types such that no face is occupied by a patch in both cubes (Fig.~\ref{fig:reducer-explain}b). The algorithm then combines the two cubes and adds an allosteric control such that all patches will begin available for binding, but binding of any patch derived from one of the cubes inactivates all patches derived from the other. The resulting allosteric cube (Fig.~\ref{fig:reducer-explain}c) can function as either one of the original two cubes, but those two functions are mutually exclusive. For an example of how this looks in practice, see the polycubes in Fig.~\ref{fig:4x4tiles}e and \ref{fig:4x4tiles}f.

In order to demonstrate this system, we ran it on a set four particle types that produces a four-by-four tile assembly. The algorithm reduced the number of particle types by one by combining the green and blue particles in the static particle set into the cyan particle (See Fig.~\ref{fig:4x4tiles}). The static and allosteric particle sets were run under the same conditions and produced yields that are about as good as the static set with larger particle set.
We also constructed a reduced swan particle set, described in Suppl.~Fig.~\ref{fig:swans}.

\subsection{Allosteric Multifarious Assembly}
One of the most appealing applications of allostery-mimetic behavior in multicomponent assembly is the realization of multifarious self-assembly. In multifarious systems, the particles can store multiple possible target assemblies, which can be selectively recalled. Our signal-passing allostery mimetic polycube designs offer a possible way of realization of multifarious behavior: a group of particle types is shared between the possible target structures. Hence, rather than having a unique set of particle types for each shape, the multifarious system can be steered such that respective particle types can appear in different target assemblies. Addition of external signals then drives the system to start self-assembling into a target structure. 

The external signals that we consider in our examples are particles of a particular species that are added to the inert mixture, which through allosteric patches activation of particles that they bind to steer the ensemble of particles to self-assemble into the desired target structure. Addition of a different groups of steering signal particles would lead the system to assemble into a different structure. We demonstrate this functionality using two rationally-designed multifarious allosteric particle sets, which we call wereflamingo (which stores 2 distinct target assemblies) and triple bridge (which has three possible target assemblies). Finally, we demonstrate a third multifarious system design, which we call stonehenge. In this case, a pre-patterned surface is prepared before simulation starts, with "pattern" corresponding to certain particle species attached to a surface. The allosteric particles that are present in the bulk can interact with the particles attached to the surface, and depending on which particle they bind to, they will start growing into a different pattern.


Allosteric multifarious assembly systems can be divided into constitutive and divergent. 
In a divergent paradigm, two or more structures share a common component, and allosteric behavior is used to control whether the common component is incorporated into one structure or the other, and to prevent the formation of chimeric structures. Wereflamingo design is an example of such a design (Fig.~\ref{fig:wereflamingo-yields}) 
In constitutive assemblies, one structure is entirely contained within another structure. Constitutive multifarious structures can be nested within each other, as in the triple bridge example (Fig.~\ref{fig:bridge-particle-sets}). 


\subsubsection{Wereflamingo multifarious set}
The wereflamingo design was created as a demonstration of divergent allosteric multifarious assembly. The design (Supp.~Fig.~\ref{fig:wereflamingo-particles-suppl}) consists of a common component containing seven particles of three types (red, yellow, and pink) which forms the "head" and "arms" of the "human" structure and the "wings" and "leg" of the "flamingo" structure. 
The particle set is designed to form entirely starting from the red particle, and contains additional allosteric logic on the red particle that makes binding to either the human torso particle (cyan) or the flamingo lower neck (dark-blue) necessary before the yellow particles can bind. The yellow particles also contain additional logic so that they must bind to the red particle before the maroon head/neck particles. The substructures which are unique to the human and flamingo contain similar logic.

\begin{figure}
    \centering
    \includegraphics[width=0.5\textwidth]{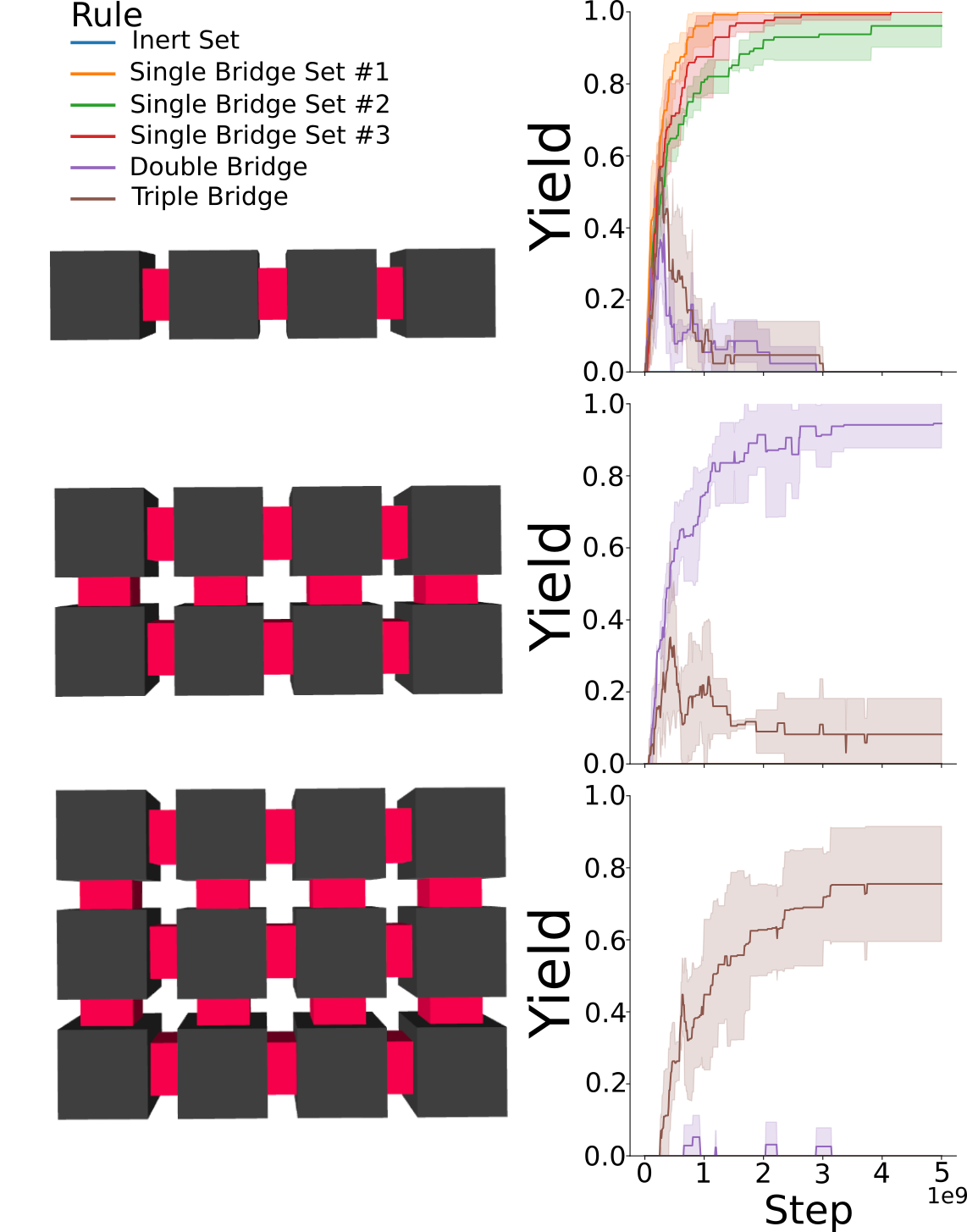}
    \caption{Yields for the triple bridge particle sets. The structures on the left correspond to the yield curves on the right. The color coding of the different particle sets corresponds to Fig.~\ref{fig:bridge-particle-sets}.}
    \label{fig:bridge-yields}
\end{figure}

The yield data (Fig.~\ref{fig:wereflamingo-yields}) show that the particle sets behave as expected. When the shared particles and those required for the human structure are present (blue line), only the human structure forms. When the shared particle types and those required for the flamingo structure are present, only the flamingo forms (orange line). When both are present, both structures form to some degree (green line). If only the shared particle set is present, no formation occurs. 




\begin{figure*}
    \centering
    \includegraphics[width=\textwidth]{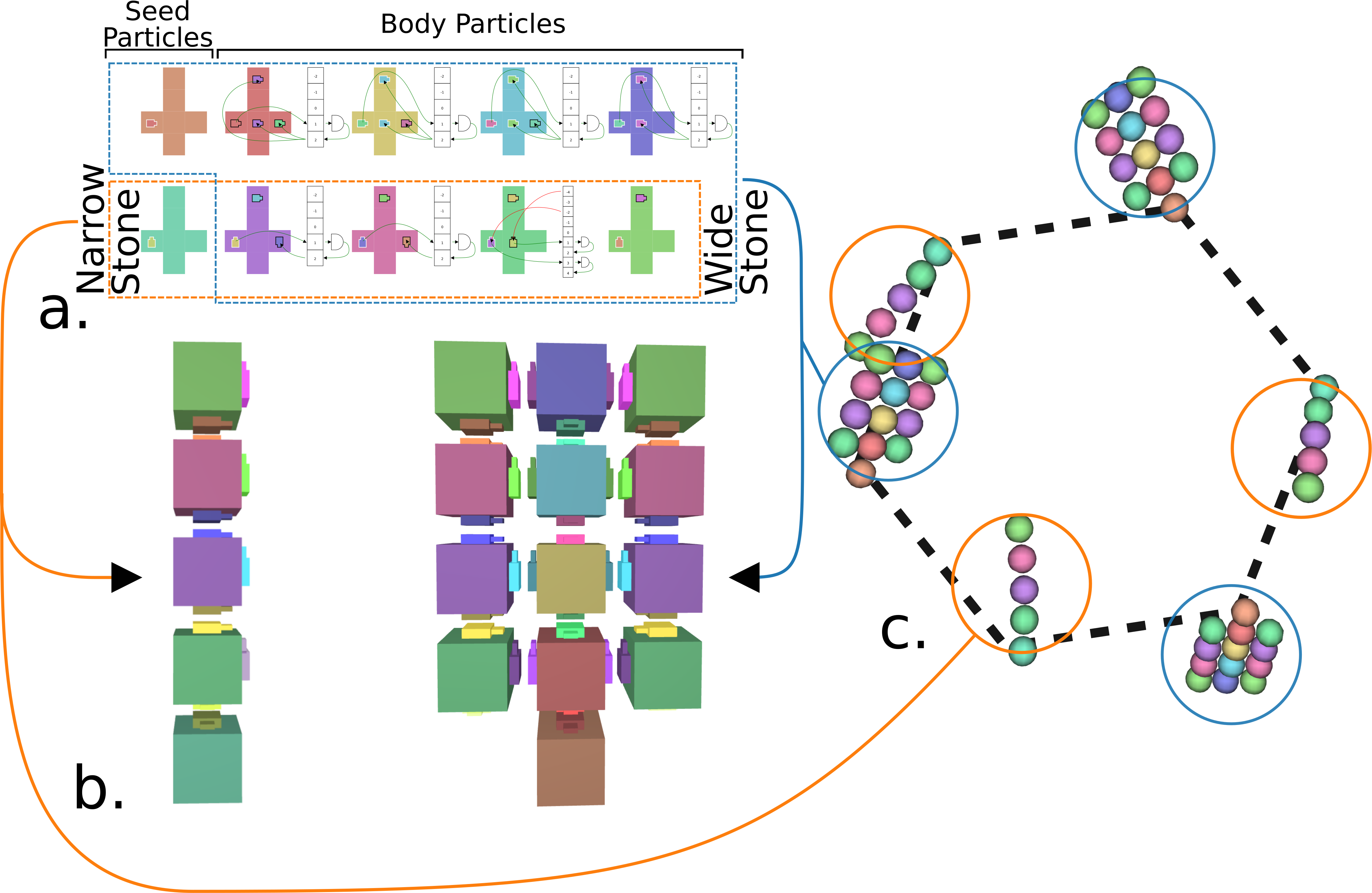}
    \caption{
        Multifarious assembly set and simulation results of the stonehenge design. {\bf a)} The particle type set that forms the stonehenge design. The particle types enclosed in the blue dashed line box are used in the large "standing stone". The particle types enclosed in the orange dashed line box are used in the small "standing stone". The leftmost two particle types are the seed particles. {\bf b)} The two "standing stone" structures visualized in Polycubes. {\bf c)} The final configuration of the stonehenge simulation, after $2.321 \times 10^{10}$ steps. The small and large "standing stones" are circled for easier identification in colors corresponding to {\bf a}. The dotted hexagon indicates the arrangement of the seed particles, which are fixed ad the vertices of the hexagon. In order to get the particle density right, the simulation  was run with $4\times$ the number of body particles required, and after the simulation was complete, particles not bound to the seeds were removed to simulate washing. The unwashed simulation is shown in Suppl.~Figs.~\ref{fig:stonehenge-blobby-nowash}~and~\ref{fig:stonehenge-pointy-nowash}.
    }
    \label{fig:stonehenge}
\end{figure*}

\subsubsection{Triple bridge multifarious set}
In order to demonstrate constitutive allosteric multifarious assembly, we designed the triple bridge rule, shown in Fig.~\ref{fig:bridge-particle-sets}. The complete particle set contains nine particle types.
The particle set was designed to self-assemble into only the single bridge (a linear tetramer of four particles), the double bridge (a pair of parallel linked linear tetramers), and the triple bridge (a triplet of parallel linked linear tetramers). The triple bridge only forms when all nine particle types are present. The single bridge is a substructure of the triple bridge consisting of the central linear tetramer. It only requires three particle types (green, red, and purple) but the allosteric rules ensure it will still form while excluding all other structures as long as the particles required to form the double and/or triple bridge are not present (see the orange, green, and red particle sets in Fig.~\ref{fig:bridge-particle-sets} and corresponding yield curve in Fig.~\ref{fig:bridge-yields}).

The double bridge can form in two ways, both using the single bridge tetramer as one of the two parallel tetramers that make the bridge "double". One formation uses the blue, lime, and orange particles while the other uses the yellow, cyan, and pink particles. The allosteric behaviors of both bridge types is identical, so for our simulations we exclusively tested the yellow, cyan, and pink version (as seen in Fig.~\ref{fig:bridge-particle-sets}). When the particles required to form the central tetramer (single bridge) and only one of the two flanking tetramers are present, the particle set will form the double-bridge topology. The results of the purple experimental group (see Fig.~\ref{fig:bridge-particle-sets}, labelled as "Double Bridge" in Fig.~\ref{fig:bridge-yields}) show  that without the blue and lime particles, the orange particle will not be able to attach and no malformed partial triple bridges will form.

\subsubsection{Stonehenge: a multifarious design from pre-patterned surface} 
We constructed the stonehenge design and particle set to demonstrate how allostery-mimetic behavior can be incorporated into surface-seeded assembly (Fig.~\ref{fig:stonehenge}a). While the stonehenge structure appears very similar to the triple bridge structure from which it was derived, it uses distinct trigger mechanism by a seeding particle attached to a surface. The key difference is in using the divergent assembly paradigm instead of the constitutive one, and in using spacial localization to overcome the challenges involved in designing divergent multifarious assembly rules. The particle sets consists of two seed particle types and eight body particle types. The seed particles were fixed in place in the simulation by a spring potential which holds them in a pre-defined position. This approach can be in practice realized e.g. by  DNA strands affixed to a surface, opening a way where e.g. dynamic process will deposit signal onto a surface, from which the desired structure will then grow. The body particles were allowed to diffuse freely.

The rule relies on the fixed seeds to prevent chimeric forms. The binding of the teal seed particle to the green particle deactivates binding of the green particle to the red particle, but the green particle has no way to communicate this information to the purple particle. In other words, the purple particle has no way to "know" whether it's part of a narrow or broad "stone" assembly. Therefore, a purple particle in a seeded "narrow stone" structure could bind to a yellow particle. By physically separating the wide and narrow stone seeds, we prevent chimeric forms by localizing the state of the yellow cube with the lateral patches (dark cyan) activated to the area around the orange seed particle, preventing it from binding to any green particles that have already bound to a narrow stone seed.
The resulting system behaves as expected in simulation, with desired structures growing from the seeded particles as shown in Fig.~\ref{fig:stonehenge}c.


\subsubsection{Possible experimental realization with DNA nanotechnology}
While we are aware that many of the logical circuits presented here strain plausibility with regards to implementation, our work has been informed by the eventual goal of realizing allosterically controlled particles {\it in vitro}. The proposed allosteric mechanism, while challenging to realize experimentally, is still an easier behavior to realize at nanoscale than other theoretically proposed "programmable matter" approaches \cite{derakhshandeh2014amoebot}. Our designs are inspired by prior work in structural and dynamic DNA nanotechnology, and we outline here a plausible scenario of realizing allosteric computation inspired by hybridization chain reaction \cite{dirks2004triggered}, based on the strand displacement mechanism in DNA nanotechnology \cite{simmel2019principles}. The DNA strand displacement (DSD) mechanism relies on a single-stranded DNA binding to an exposed single-stranded region on a pre-formed DNA duplex, and removing (displacing) the incumbent strand from the duplex.

\begin{figure}
    \centering
    \includegraphics[width=0.5\textwidth]{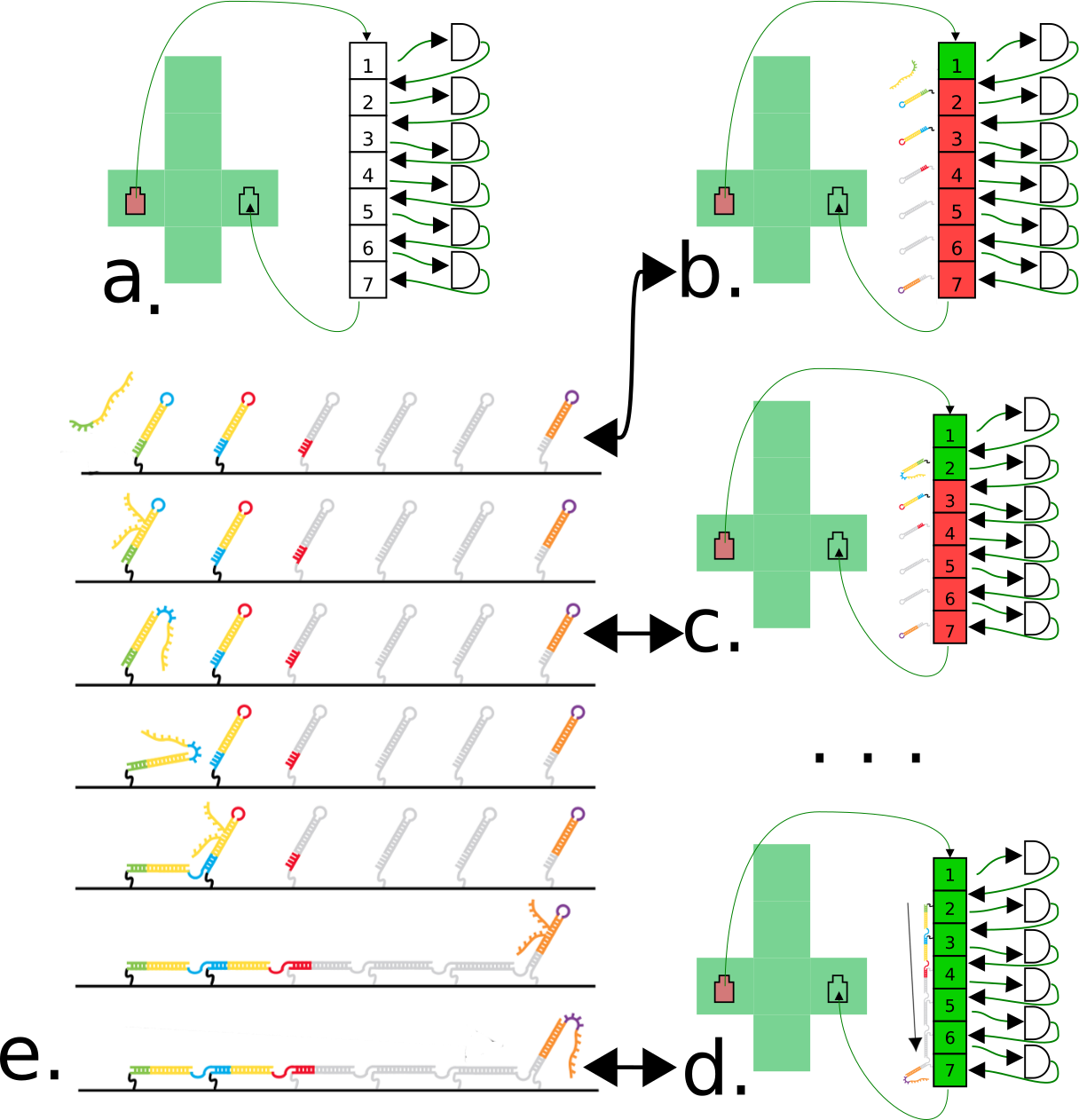}
    \caption{A DNA hybridization chain reaction could correspond to our model of allosteric particles. {\bf a)} A particle type schematic in which a an allosteric circuit propagates a signal from the red patch to the green patch. {\bf b)} The schematic from {\bf a}, juxtaposed with a proposed DNA strand displacement cascade. At this point in the cascade, binding to the red patch has released a ssDNA molecule (top of state view). We've represented the presence of this molecule (assumed to be tethered, although that isn't shown here) the box labelled "1" {\bf c)} The diagram from {\bf b}, but now the first hairpin has been opened, , which we represent by turning the box background from red to green. {\bf d)} The diagram from {\bf b} and {\bf c}, once the DNA strand displacement cascade has arrived at the last box, activating the green patch. {\bf e)} A more complete visualization of the strand displacement cascade.}
    \label{fig:patchy-to-dna}
\end{figure}

Previous work adapting patchy particle models to DNA nanostructures has used DNA wireframe origami with single-stranded DNA overhangs as "sticky ends" (patches) \cite{tian2016lattice,liu2023inverse}. Complementary DNA strands then represent compatible colors. 
Additionally, the DNA strand displacement reactions can be also been shown to work where the interacting strands are attached to a surface, such as flat DNA origami sheet \cite{seelig_enzyme-free_2006, dalchau_probabilistic_2015, thubagere_cargo-sorting_2017, chatterjee_spatially_2017}. In the strand displacement paradigm, the displaced strand can then go on to engage in a further DSD reaction, creating a signal passing cascade.

Inspired by the successfully realized surface-based strand displacement systems, our proposed method of implementation of allosteric communication from one binding site of the DNA nanoparticle would be to attach to activator patches a complementary strand that forms a hairpin. The hairpin stem would be displaced by the fully-complementary sticky end on the corresponding patch and trigger a DSD cascade, eventually removing or introducing a blocking strand on the patch to be activated or deactivated. Part of this mechanism is shown in Fig.~\ref{fig:patchy-to-dna}. Each box to the right side of the schematic Fig. \ref{fig:patchy-to-dna}a-d represents a DNA hairpin, with the boxes linked by operations representing the action of a hairpin opening by a DSD reaction. When the red patch is bound, the hairpin stem is displaced and opened, represented by the box turning from red to green (Fig.~\ref{fig:patchy-to-dna}b). An DSD reaction (represented by the operation with the hairpin box 1 as an input) then opens the hairpin represented by box 2, shown by box 2 going from red to green (Fig.~\ref{fig:patchy-to-dna}c.). A cascade then occurs with each hairpin opening the subsequent hairpin, culminating in the hairpin at box 7 (Fig.~\ref{fig:patchy-to-dna}d), which corresponds to the activation of the green patch.

Reconfigurability of DNA nanostructures \cite{andersen_self-assembly_2009} and DSD cascades localized to the surfaces of DNA nanostructures \cite{chatterjee_spatially_2017} can hence serve as a starting point for experimental realization of our allosteric model presented in this work. Even for a signal-passing mechanism which doesn't contain logic, realization of this could open up new possibilities for the field of multi-component self-assembly, as shown by our pyramid, triple bridge, and stonehenge simulations.

The polycube model we have outlined in this article has several limitations: i) particles have cubic arrangement of patches and at most one patch on each cube face, ii) patch binding is also irreversible. The second limitation was deliberately imposed to mimic the irreversible nature of DNA strand displacement reactions, which we envisioned as the potential mechanism for in vitro implementation of allostery-mimetic behavior in DNA nanoparticles. The first limitation was imposed to simplify the range of simulations we examine, and are not based in any physical constraint. If patches are envisioned as DNA sticky ends, it is realistic to include multiple sticky ends with different sequences on a single face (indeed, this is how we implement torsional modulation in vitro), and there it's reasonable that inactivation of a patch could detach a particle instead of preventing it from binding. We are also not restricted to just cubic shape, as many different 3D DNA origami shapes have been designed and experimentally realized previously \cite{gerling2015dynamic,zhang2015complex}. 


\section{Conclusion}

Here, we have introduced an allostery-mimetic model for design of self-assembling structures. While most previously developed models of multicomponent self-assembly assumed that the interaction sites between the particles are static (i.e. always available for binding), we allow some of our particles to change their binding site availability  based on other interactions in which a particle was or is involved. We have shown with patchy particle molecular dynamics simulations that this ability allows for improvements to existing self-assembly designs, in particular i) it can improve the yield by helping avoid misassambled states observed in static systems; ii) can reduce the total number of different particles needed to assemble target shape. Even more excitingly, our model is able to realize multifarious assemblies, where a set of particles is inert until an external triggers signal is introduced, which causes the system to start self-assembling into a shape based on the type of the trigger provided. We showed different kinds of the trigger examples: in once scenario, a set of particles is introduced into the system, and in second scenario, a prepatterned surface is used to direct formation of different structures in different locations. These structures grow from the shared set of particles in the bulk, and it is just the provided surface-attached seed that determines the structure formed.  

Our model has been specifically developed with the intent that it could be eventually realized for 3D assemblies with DNA nanotechnology. We outlined a plausible implementation of some of the behavior by using DNA nanostrucutres with surface-bound DNA strand displacement cascade.

In the future, our model can be extended to further allow for more realistic representation of the experimental design, and account e.g. for leaky reactions of DNA strand displacement by - for example - having probability of allostery function and incorrect binding, with the goal to design more robust systems. This model combined with the experimental realization with DNA nanostructure could allow us to use nucleotide-level coarse-grained simulations to parameterize state transition probabilities, and develop multiscale models that will guide the experimental work. 


A second promising avenue of research is to explore the possibility of using allosteric behavior to break bonds between particles. This line of work could allow us to make "disassembleable" structures which form under certain conditions and then can be triggered to disassemble by an external signal, thus enabling realization reconfigurable assemblies that assemble and disassemble.

Overall, this pipeline can guide new designs for application in nanomanufacturing. For example, a dynamic process will produce a specific pattern on the surface, that will then grow different bridge connections from a multifarious set of allosteric particles. Applications can include an artificial cell mimicking microtubule growth. In other possible applications, an external signal (e.g. RNA or DNA strand of interest) will trigger the particle swarm to assemble into a particular functional complex. Our future work will focus on proof-of-principle experimental realization of the simpler examples the allostery-mimetic behavior.


\section*{Acknowledgements}
This result is part of a project that has received funding from the European Research Council (ERC) under the European Union’s Horizon 2020 research and innovation programme (Grant agreement No. 101040035).


\clearpage
\newpage

\onecolumngrid
\begin{center}
\textbf{\large Supplementary Materials}
\end{center}

\setcounter{figure}{0}
 \makeatletter 
 \renewcommand{\thefigure}{S\@arabic\c@figure}
 \setcounter{equation}{0}
 \renewcommand{\theequation}{S\@arabic\c@equation}
 \setcounter{table}{0}
 \renewcommand{\thetable}{S\@arabic\c@table}
  \setcounter{section}{0}
 \renewcommand{\thesection}{S\@arabic\c@section}
  \renewcommand{\thesubsection}{S\@arabic\c@section.\@arabic\c@subsection}
   \renewcommand{\thesubsubsection}{S\@arabic\c@section.\@arabic\c@subsection.\@arabic\c@subsubsection}

   \section{Additional Information}
   \subsection{A Detailed Explanation of States and Operations}
   Each particle's state $s$ is described by a series of Boolean variables $s_i$, all of which start as $F$ except variable $s_0$, which starts as $T$. Variable activation is irreversible in our model, so once a state variable is flipped from $F$ to $T$, it cannot be flipped back. In addition to the "real" state variables, a particle's state also includes a set of "virtual" state variables defined so that the virtual (negative) of a variable always has the value of the boolean negation of the real variable (so value of $s_{-i} \equiv \neg s_i$). Virtual state variables are not assignable but otherwise function the same as real state variables. All patches are assigned a controlled variable and an activation variable. A patch is active if and only if controlling state variable is $T$. When a patch forms a bond with another patch, the patch's activation variable is set to $T$. 
   In our schematics, we show a particle's state variables as white squares on the right of a particle, with the variable number within each square and arrows to and from the square showing how it is set by or sets patches.
   
   The second component of our allosteric polycube model is what we call "operations". An operation consists of a set of state variables that serve as the inputs, and a state variable which serves as the output. Input variables can be real or virtual, but as state variables are irreversible the output variable must always be real. An operation "fires" when all its input variables are set to $T$. The number of input variables can be as low as 1 and has no limit imposed by the model, although no example in this paper has an operation with more than two inputs. While operations function as and-gates, logical or-gates can be constructed by targeting multiple operations at the same variable (see Fig.~2d). 
   
   \subsection{The Rationale Behind Piecewise Assembly}
\begin{figure}
       \includegraphics[width=\textwidth]{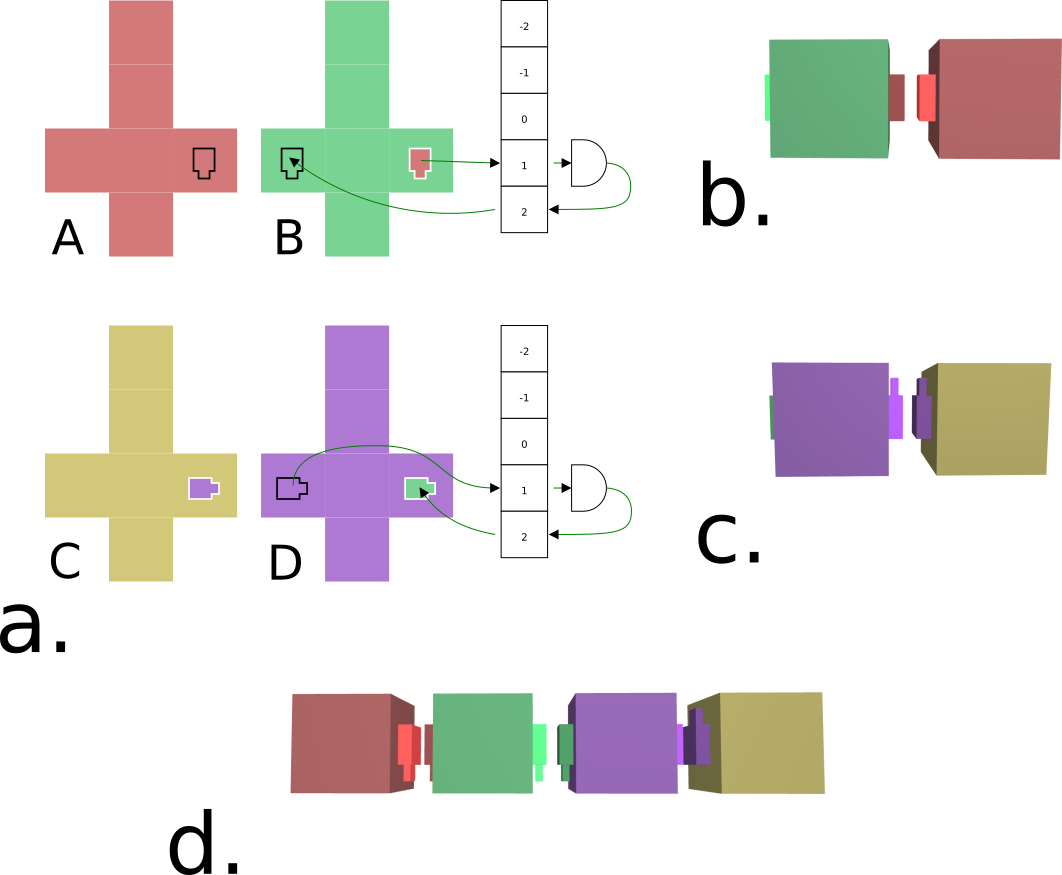}
       \caption{The non-commutativity of allosteric multicomponent assembly, as expressed symbolically in Eq.~\ref{eqn:noncommutivity}. {\bf a)} Particle types $A$, $B$, $C$, and $D$. The green patches on particles types $B$ and $D$ are allosterically activated by the red and purple patches on their respective particles. {\bf b) \& c)} The structures formed by the stochastic assembler when piecewise assembly is turned off. {\bf d)} The structure formed by the stochastic assembler when piecewise assembly is turned on.}
       \label{fig:piecewise}s
   \end{figure}

   Piecewise assembly is necessary to circumvent an issue with the stochastic assembler which did not exist in the non-allosteric model. This behavior arises because the binding behavior of allosteric particles does not follow an associative identity property. In other words, the order in which binding events happen matters. To understand why this is the case, it's helpful to imagine the formation of a simple tetrameric structure $ABCD$ (Fig. \ref{fig:piecewise}d), where A, B, C, and D are shown in Fig. \ref{fig:piecewise}a. In this particle set, the allosteric controls on particles $B$ and $C$ give rise to:
   \begin{align}
       A + B + C + D \neq (A + B) + (C + D)
       \label{eqn:noncommutivity}
   \end{align}

   Once $A$ and $B$ are bound together, an attempt to add a single particle of type $D$ will fail because - without having already formed an attachment with particle type $C$, the green patch on $C$ (required to bind to $B$) is not active. The same problem arises no matter which particle in this set we choose as our starting point; there is no way to construct the full structure by adding one particle at a time. We see this in the stochastic assembler; when run on the $ABCD$ structure without piecewise assembly, it produces the dimers in Fig. \ref{fig:piecewise}b and \ref{fig:piecewise}c, but never the tetramer. When we activate piecewise assembly (with piece size 2 or higher), the tetramer forms correctly in the stochastic assembler (Fig. \ref{fig:piecewise}d).
   This is however not an issue for the patchy particle molecular dynamics simulations (see below), which intrinsically allow for piecewise assembly.
   
   \subsection{Torque Potential Modulation}
   In section~\ref{sec:patchy-methods} of the main text, we describe a set of torque potential patchy interaction parameters $\Delta$, $\Delta_c$, $a$, and $b$ which were used in the equations for the patch-patch energies. All simulations in the main text used the patchy parameter values described therein. However, we also ran simulations with other sets of $\Delta$, $\Delta_c$, $a$, and $b$ values. Following the convention used in [19], we enumerated the torque parameters as narrow types. All simulation in the main text used narrow type 1. Higher narrow types correspond to tighter directional constraint.
   
   \begin{table}
       \centering
       \begin{tabular}{c|cccc}
           Narrow Type & $\Delta$ & $\Delta_c$ & $a$ & $b$ \\
           \hline
           \hline
            0 & 0.7 & 3.105590 & 0.46 & 0.133855 \\
            1 & 0.2555 & 1.304613 & 3 & 0.730604 \\
            2 & 0.2555 & 5 & 2.42282 \\
            3 & 0.17555 & 0.438183 & 13 & 8.689492
       \end{tabular}
       \caption{Values of $\Delta$, $\Delta_c$, $a$, and $b$ for narrow types 0, 1, 2, and 3.}
       \label{tab:narrow-types}
   \end{table}
   
   \subsection{Molecular Dynamics Simulation Conditions}
   \subsubsection{Simulations of 4x4 Tiles}
   4x4 tile simulations were run with 4 and 8 assemblies (4-assembly data is in Fig.~\ref{fig:tile4x4-suppl}). Each assembly of the static particle set contained four each of the red, yellow, green, and purple particle types. Each assembly of the reduced particle set contained four each of the red and yellow particle types and eight of the cyan particle type. Simulations were run at temperatures T=$0.01$, $0.02$, $0.05$, $T=0.1$. Four replications were run with each set of simulation conditions. 
   
   \subsubsection{Simulations of pyramids}
   Each pyramid assembly contains 18 particles of five species, and all simulations were run with enough particles for 8 assemblies Each assembly contained one yellow particle, four each of purple, red, and green particles, and five cyan particles. All pyramids were run at temperatures T=0.01, T=0.025, and T=0.05. Simulations were run for $5 \times 10^9$ steps, with cluster data recorded every $10^7$ steps. Four replications were run for each set of simulation conditions.
   
   \subsubsection{Simulations of Solid Cubes}
   Simulations of solid cube assemblies were performed with 12 structures worth of particles, where each structure had 1 yellow particle, 12 red particles, 8 green particles, and 6 purple particles. Simulations were run at temperatures of T=0.01 and T=0.05. Five replicas were run of each set of simulation conditions. Simulations were run for $2.5 \times 10^9$ steps. Cluster topologies were recorded every $10^7$ steps.
   
   \subsubsection{Simulations of Wereflamingos}
   Simulations of wereflamingos were performed with 8 total structures worth of particles in each simulation. The exact number of particles varied between the simulations for human setup, flamingo setup, mix setup, and neither setup. The human structure contains a red particle, 3 yellow particles, 2 purple particles. 5 pink particles, a cyan particle, and a green particle. The flamingo structure contains a red particle, 3 yellow particles, 5 pink particles, and one each of lime, blue, and purple particles. The human and flamingo setups had 8 of their respective assemblies. The simulation intended to produce both humans and flamingos has 4 of each assembly. The simulation intended to produce neither humans nor flamingos contained 4 of each assembly, sans the blue and cyan particle types. particles total. The simulations were run at T=0.01, T=0.02, and T=0.05. We ran four replicas for each set of simulation conditions.

   \subsubsection{Simulations of Triple Bridges}
   We ran simulations of six different subsets of the bridge particle set (See Fig. 10). In all simulations, we provided enough particles to construct 8 assemblies. A complete triple bridge assembly contained two each of the orange, purple, and yellow particle types and one each of the other six types. All particle sets were subsets of this particle set, where particle types were included or not included but the number of each particle type included was kept constant per assembly unit. The particle sets we tested are shown in Fig. 10. In the particle subsets that deliberately included excess particle types (blue, green, and red in Fig. 10) we treated one assembly worth of particles to be the number of particles that would be used to construct a triple bridge, if all required particle types were present. We tested each particle group at temperatures T=0.01, T=0.02, and T=0.1, and ran four replicas of each set of simulation conditions.

   \subsubsection{Simulations of Stonehenge Designs}
   The stonehenge simulation was run in a simulation box of $24 \times 24 \times 7$ simulation units, with periodic boundary conditions on the 24-unit dimensions. Each "narrow stone" structure contained one each of the lime, pink, green, and purple particles, not counting the seed. Each "wide stone" structure contained two "narrow stones" worth of particles plus one each of the blue, cyan, yellow, and red particles, not counting the seed. In order to reach the required particle density for self-assembly to occur in a reasonable timeframe, we introduced $4\times$ the number of body particles required for the assemblies in addition to the six seed particles, a total of $198$ particles. This produced a particle density of $\approx0.05$. The simulation was run for $2.321\times10^{10}$ steps.

   \subsection{Analysis Cutoff Points}
   We have found that we sometimes observe interesting trends when we analyze the same data with different analysis parameters. In addition to testing the same data for formation of different structures (see sections in the main text on multifarious self-assembly, Figs.8-11), we explored the effects of modulating the analysis cutoff point on computed yield. The analysis cutoff point is a parameter in our analysis algorithm which requires that clusters have a minimum level of formation for them to be included in subsequent yield calculations. If the ratio of the size of the cluster to the size of the target structure is less than the cutoff, the yield for that cluster is treated as $0$. Otherwise, the yield for the cluster is the aforementioned fraction.
   
   The effects of this are most notable in our yield-improving structures. For the pyramid, at a cutoff point of 0\% (no requirement of structure size, bottom rows of Figs.~\ref{fig:pyramid-wt-comprehensive}, \ref{fig:pyramid-x1-comprehensive}, \ref{fig:pyramid-x4-comprehensive}), we see very little impact on the yield curves. However at 100\% cutoff, where every particle needs to be present for it to be counted as producing any yield, the effects of allosteric control are even more dramatic (compare Figs.~\ref{fig:pyramid-wt-comprehensive} and \ref{fig:pyramid-x1-comprehensive}, top rows).
   
   \section{Figures}
   
   \begin{figure}
       \centering
       \includegraphics[width=\textwidth]{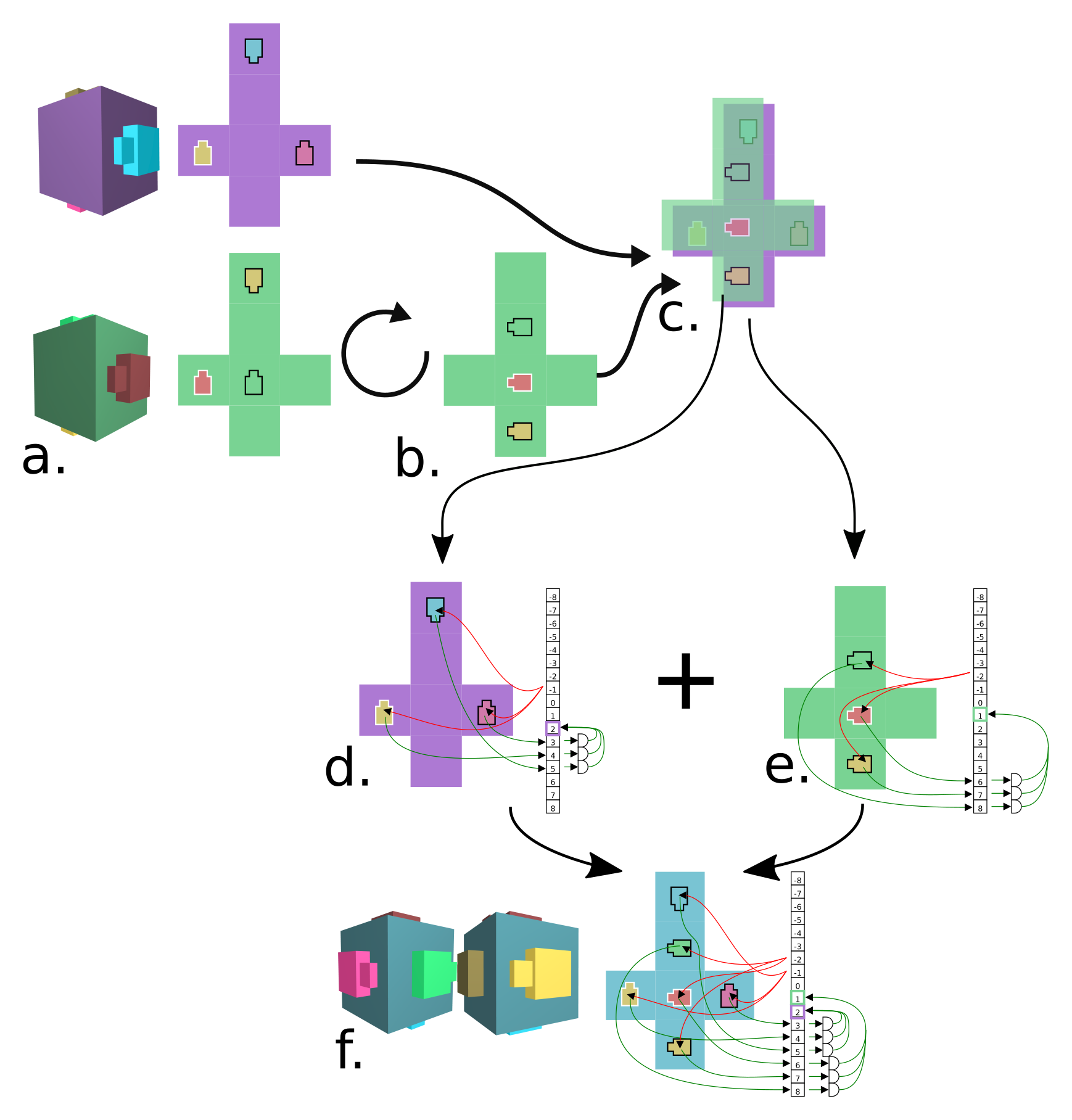}
       \caption{Expanded version of Fig.~7, showing the algorithm in more detail. {\bf a)} The two particle types to be merged. {\bf b)} The green particle type is rotated so that it can be superimposed on the purple particle type without any two patches sharing a face. {\bf c)} The green particle type, shown overlaid on the purple particle type. {\bf d \& e)} Allosteric control is added to the green and purple particles. {\bf f)} The particles, with allostery added in {\bf e} and {\bf f}, are merged to produce the cyan particle type.}
       \label{fig:reducer-suppl}
   \end{figure}
   \begin{figure*}
       \centering
       \includegraphics[width=\textwidth]{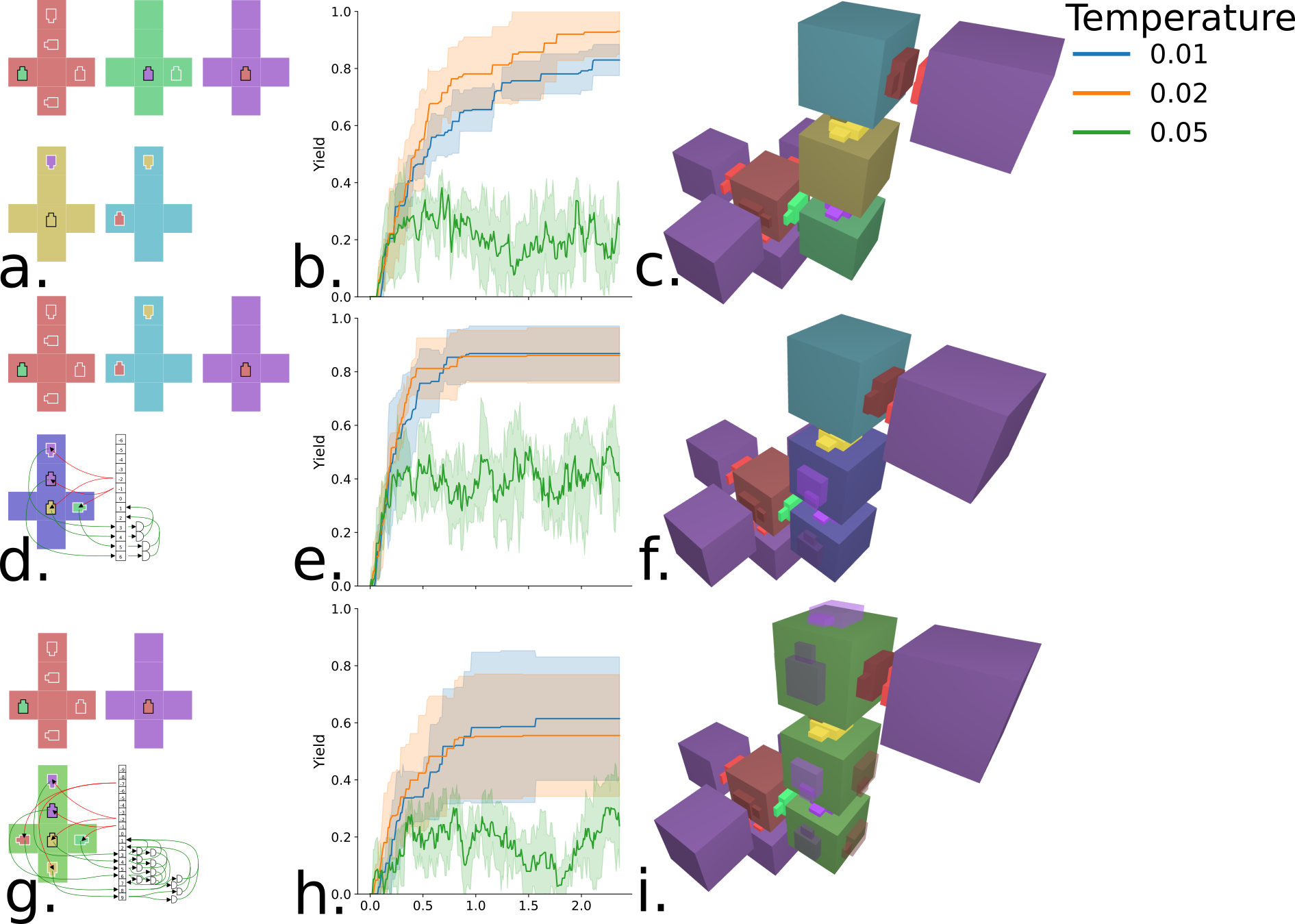}
       \caption{Static and reduced swan designs and yield curves. {\bf a)} The static particle set. {\bf b)} The yield curve for the static particle set. {\bf c)} A polycube of the structure assembled from the static particle set. {\bf d)} The best reduced particle set, where the green and yellow particle set in {\bf a} are merged using the merge algorithm (See Fig.~\ref{fig:reducer-suppl}). {\bf e)} The yield curve for the particle set in {\bf d}, which does not have noticeably different final yield than the static particle set. {\bf f)} The polycube formed by the particles in {\bf d}. {\bf g)} A reduced particle set where the cyan, yellow, and green particle types in {\bf a} are merged using the merge algorithm. {\bf h)} Yield curve for the particle set in {\bf g}. This particle set does not produce good yield because the concentrations of particle states that are kinetically favored are not the concentrations required for high yield. {\bf i)} The polycube formed by the particle set in {\bf g}.}
       \label{fig:swans}
   \end{figure*}
   
   \begin{figure*}
       \centering
       \includegraphics[width=\textwidth]{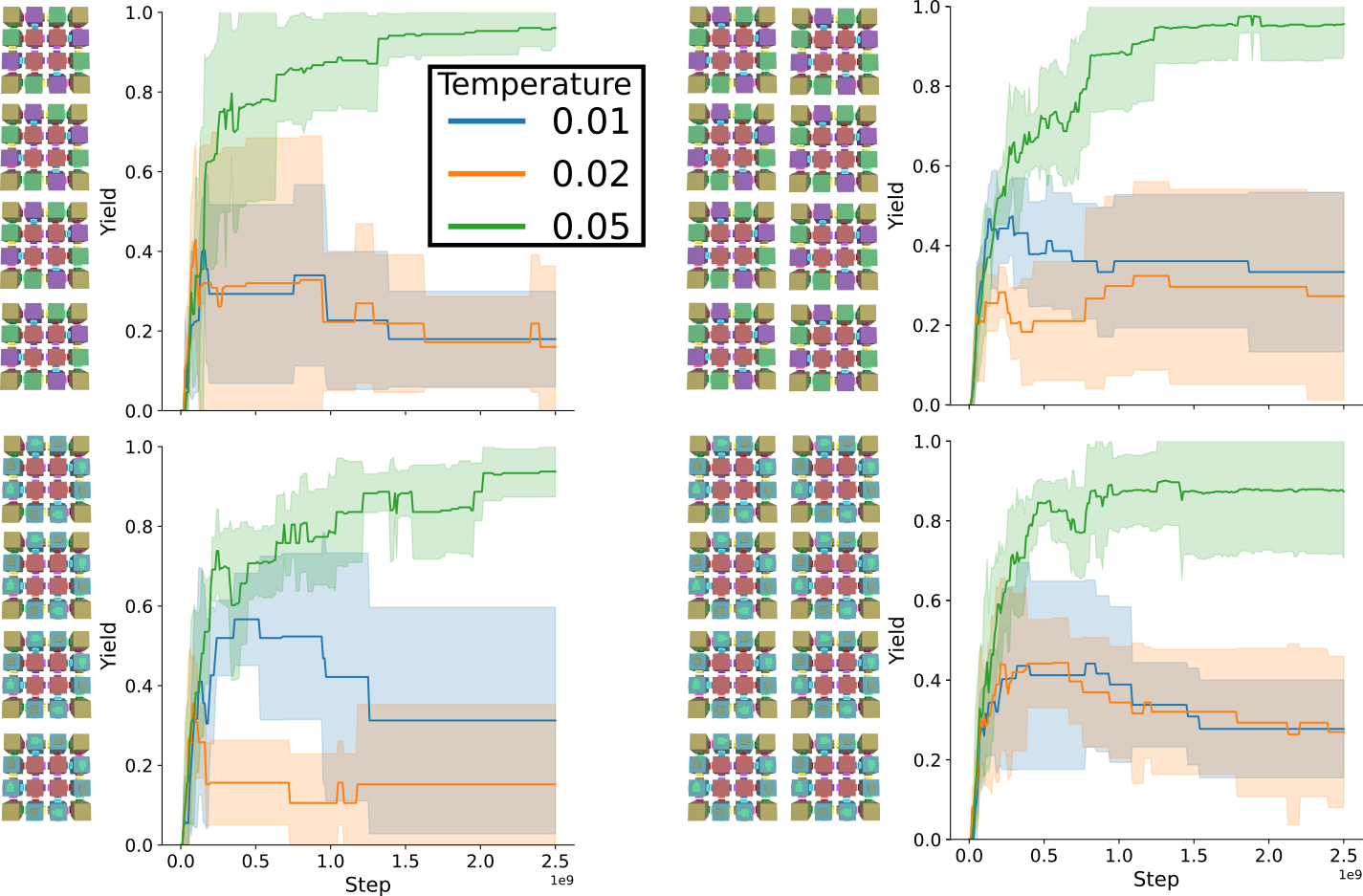}
       \caption{Data for simulations of the 4x4 tile (particle set shown in Fig.~6) with varying assembly counts. The simulations on the right were run with four tiles worth of particles, the simulations on the left with eight. The top row simulations are of the static particle set, the bottom with the allosteric particle set. }
       \label{fig:tile4x4-suppl}
   \end{figure*}
   
   \begin{figure*}
       \centering
       \includegraphics[width=.6\textwidth]{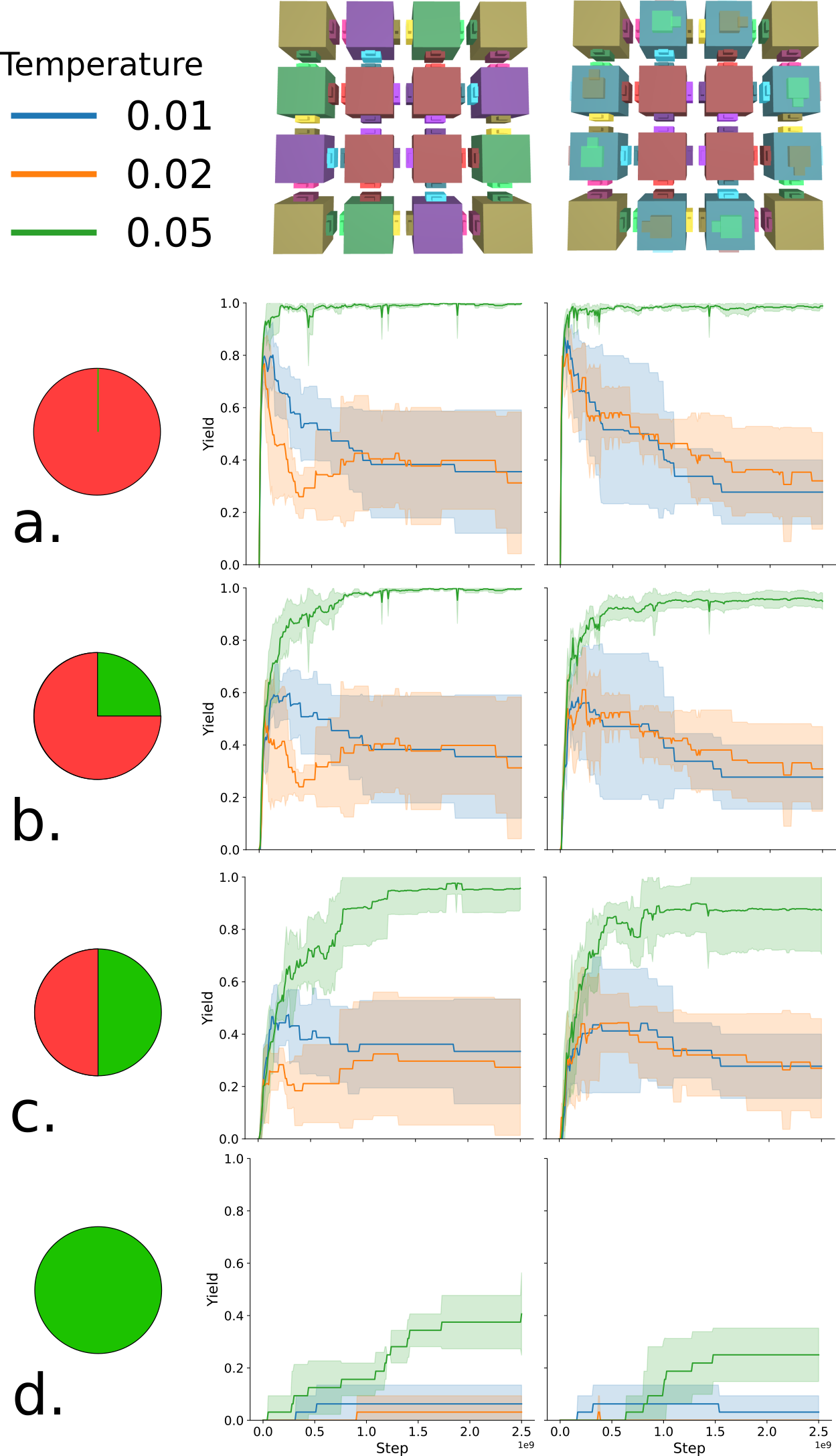}
       \caption{4x4 tile yield curves with varying cutoff point. Analysis at different yield curves demonstrates that there are no kinetic traps in the design that would, for example, halt assembly at 80\%. The left column is the static tile particle set, the right column is allosteric. {\bf a)} 4x4 tile yield curves at 100\% cutoff. {\bf b)} 4x4 tile yield curves at 75\% cutoff. {\bf c)} 4x4 tile yields at 50\% cutoff. {\bf d)} 4x4 tile yields with 0\% cutoff (no size-filtering).}
       \label{fig:tile4x4-cutoffs}
   \end{figure*}

   \begin{figure*}
       \includegraphics[width=\textwidth]{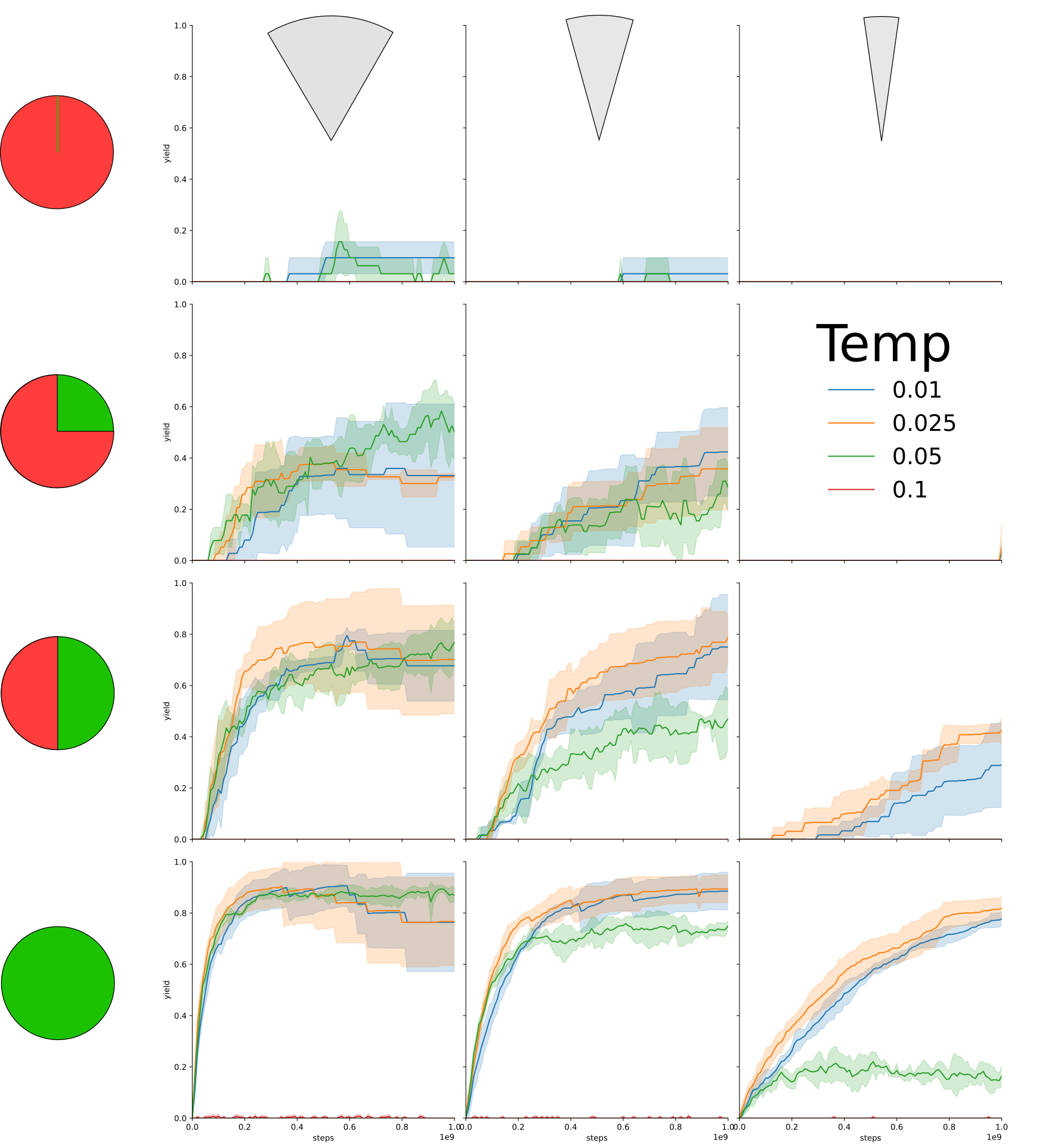}
       \caption{Static (Version 1, see Fig. 4) pyramid design yield curves across narrow type and yield. The columns are different simulations that differ by narrow type (narrow types 1, 2, and 3). The rows are different cutoff levels for analysis (100\%, 75\%, 50\%, and 0\%, represented by the circles); all plots in a given column are the same simulation data, processed with different analysis parameters.}
       \label{fig:pyramid-wt-comprehensive}
   \end{figure*}
   
   \begin{figure*}
       \includegraphics[width=\textwidth]{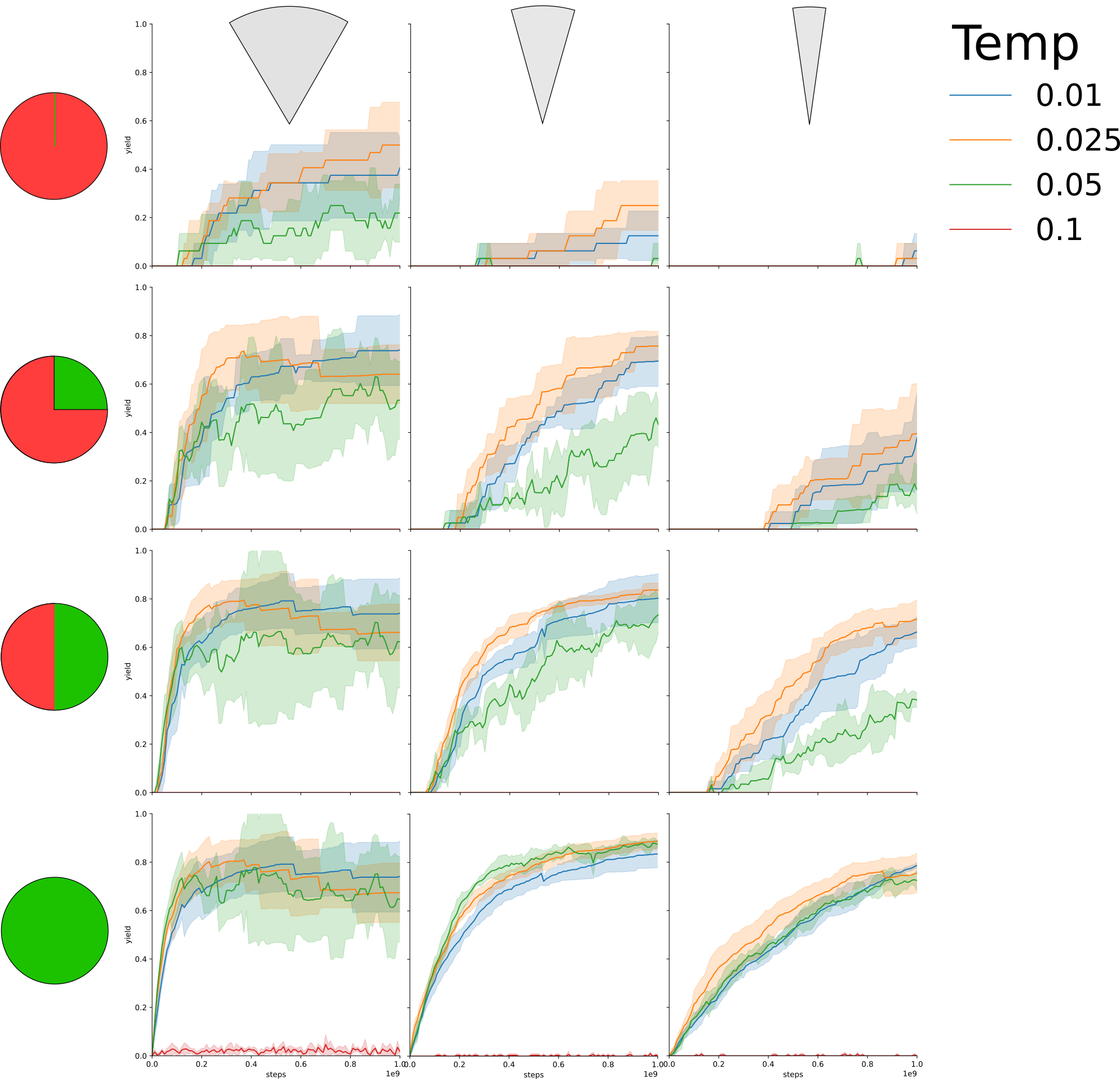}
       \caption{Version 2 (see Fig. 4) pyramid design yield curves across narrow type and yield. The columns are different simulations that differ by narrow type (narrow types 1, 2, and 3). The rows are different cutoff levels for analysis (100\%, 75\%, 50\%, and 0\%, represented by the circles); all plots in a given column are the same simulation data, processed with different analysis parameters.}
       \label{fig:pyramid-x1-comprehensive}
   \end{figure*}
   
   \begin{figure*}
       \includegraphics[width=\textwidth]{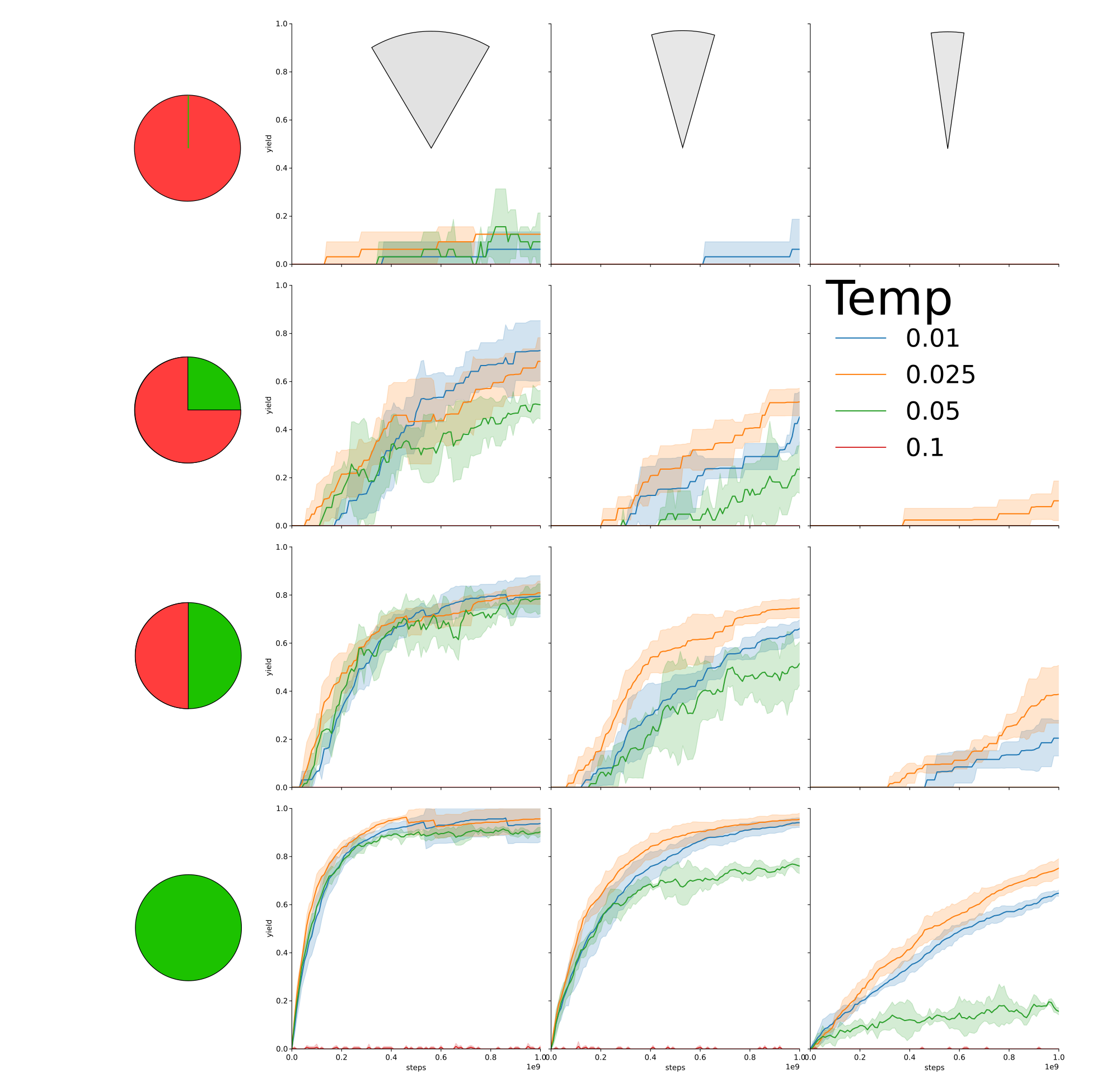}
       \caption{Version 3 (see Fig. 4) pyramid design yield curves across narrow type and yield. The columns are different simulations that differ by narrow type (narrow types 1, 2, and 3). The rows are different cutoff levels for analysis(100\%, 75\%, 50\%, and 0\%, represented by the circles); all plots in a given column are the same simulation data, processed with different analysis parameters.}
       \label{fig:pyramid-x4-comprehensive}
   \end{figure*}
   
   \begin{figure*}
       \centering
       \includegraphics[width=\textwidth]{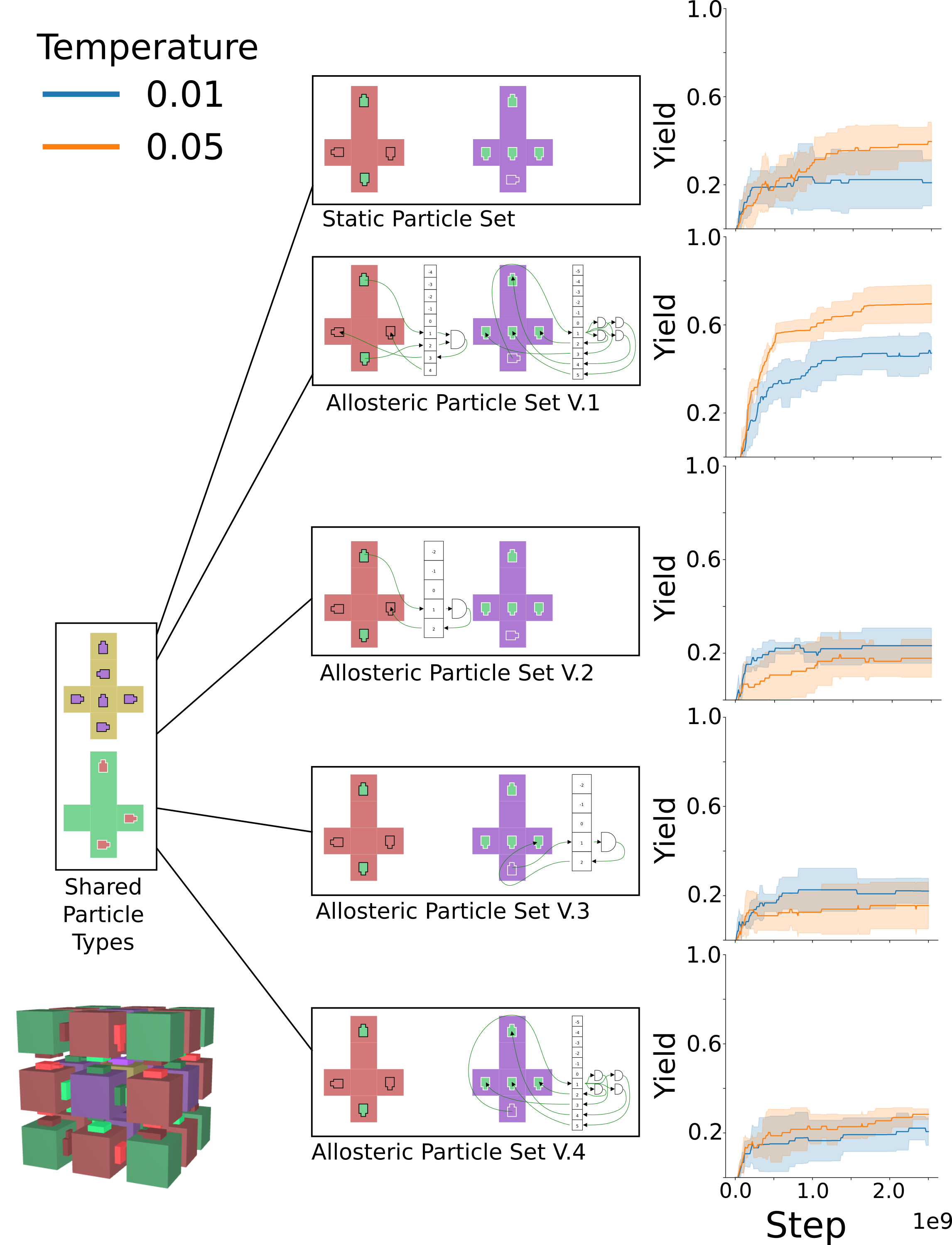}
       \caption{Comparison of yields of static solid cube with all our allosteric particle sets. Of the four allosteric particle sets, only V.1 produced an improvement in yield. The static vs. V1 yield curves are also shown in Fig.~5.}
       \label{fig:solid-cubes-suppl}
   \end{figure*}

   \begin{figure*}
       \centering
       \includegraphics[width=\textwidth]{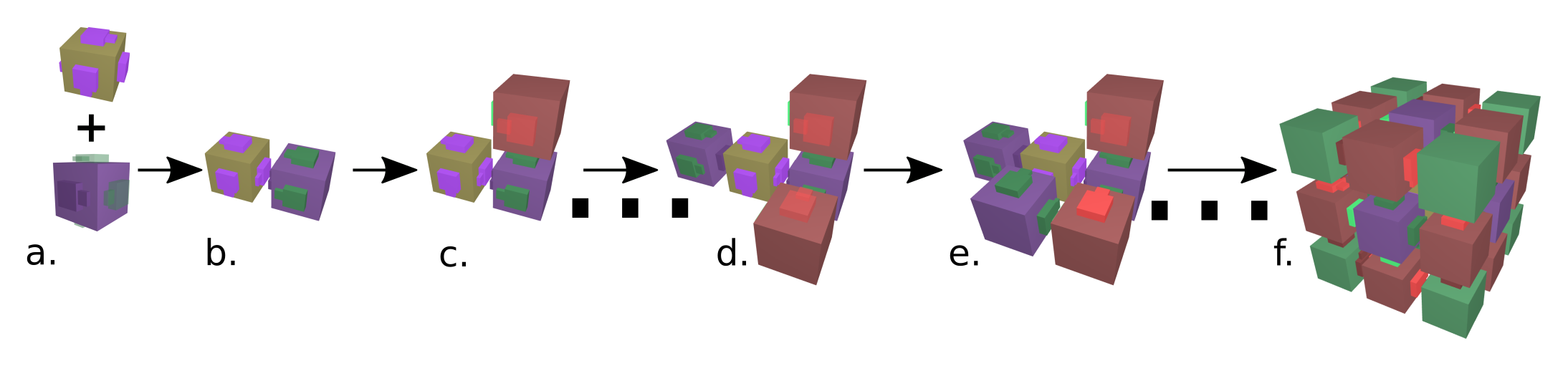}
       \caption{The assembly of the allosteric cube (Version 1, cf. Fig.~\ref{fig:solid-cubes-suppl}) in the polycubes (stochastic lattice) model. {\bf a)} The yellow and purple particle types bind to begin assembly. The green patches on the purple particle aren't active (indicated by transparency) until the purple patch is bound. {\bf b)} Binding to the yellow particle activates the green patches on the purple particle. {\bf c)} Now that the green patches on the purple particle are active, the red particle is able to bind. The red patches on the red particle are still inactive because they require binding to both green patches to activate. {\bf d)} A few more particles are added by the stochastic assembly. These intermediates are not shown. {\bf e)} Binding of a purple particle to the yellow particle adjacent to one of the red particles causes the red particle's second green patch to form a bond, which activates the red patches. {\bf f)} After more particles are added, the cube is fully assembled.}
       \label{fig:solid-cube-assembly}
   \end{figure*}
   
   \begin{figure*}
       \centering
       \includegraphics[width=\textwidth]{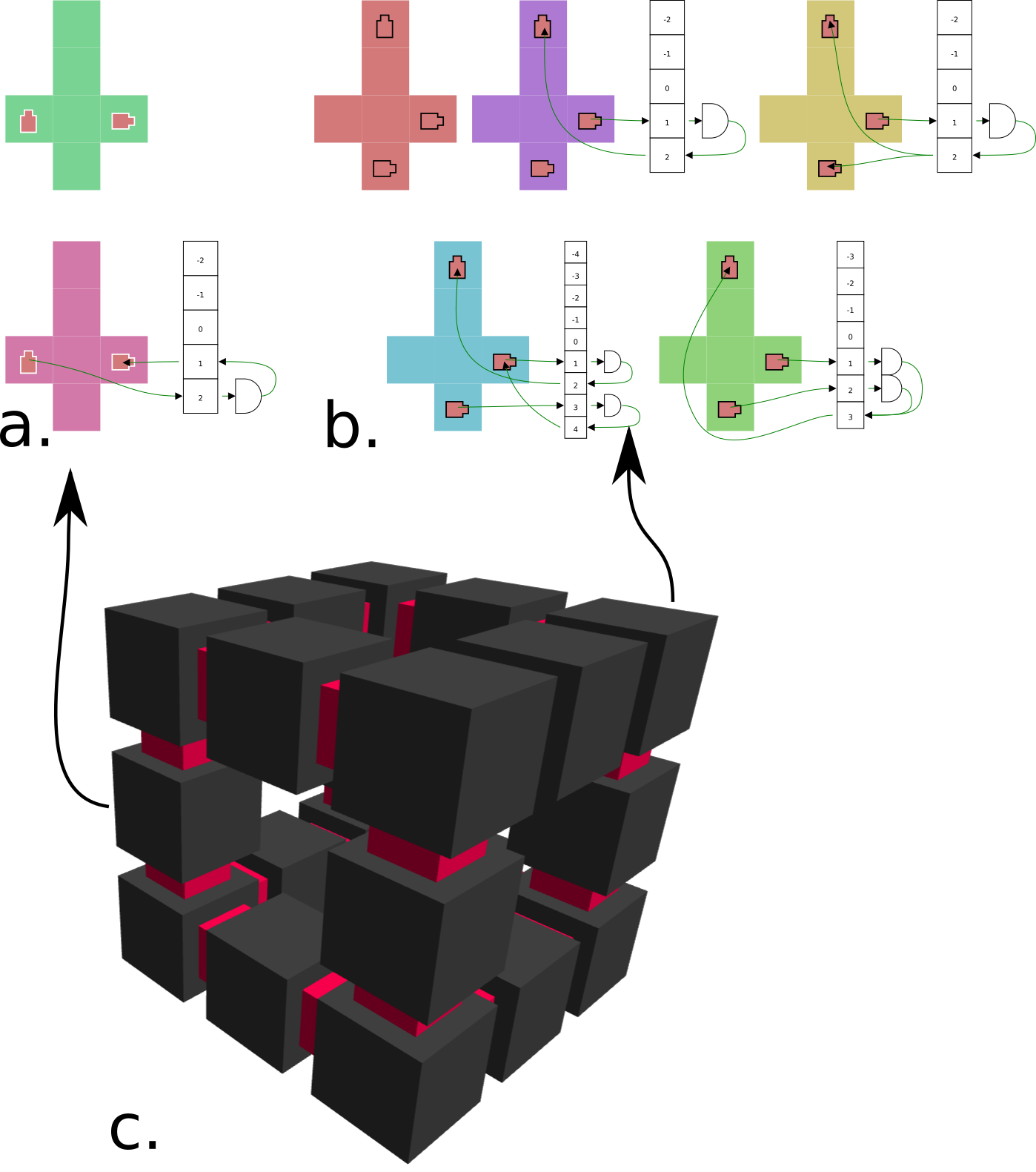}
       \caption{Particle types used to test wireframe cube particle set. {\bf a)} The two possible particle types for the wireframe cube edge. {\bf b)} The five possible particle types for the wireframe cube corner. The cyan particle type approximates and-gate behavior. {\bf c)} The wireframe cube structure, with arrows showing which parts of the structure correspond to (a) and (b).}
       \label{fig:wfc-particle-types}
   \end{figure*}
   
   \begin{figure*}
       \centering
       \includegraphics[width=\textwidth]{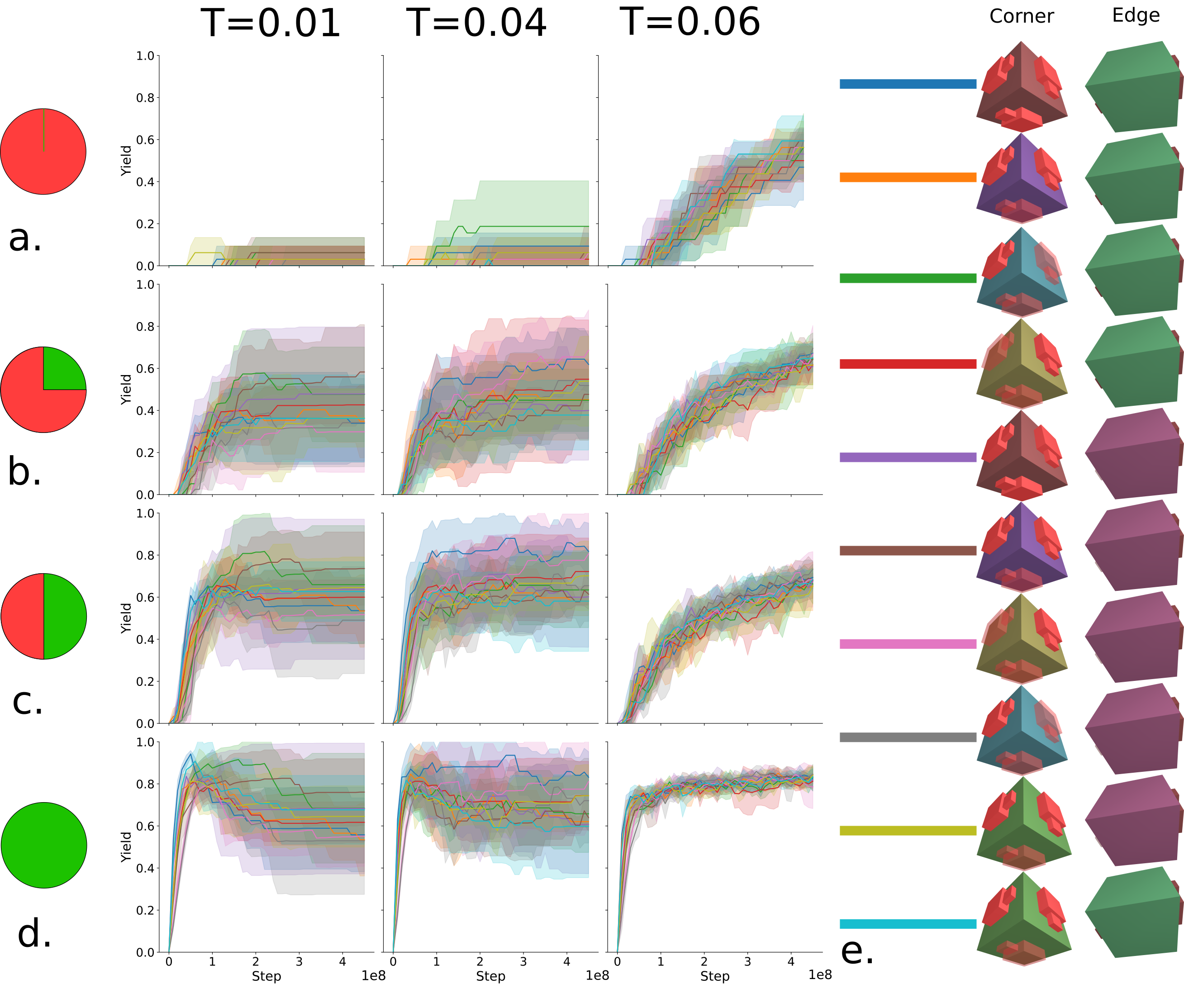}
       \caption{Yield curves for allosteric wireframe cubes. We designed a very large variety of possible allosteric versions of the two particle types in minimal wireframe cube particle set (See Fig.~\ref{fig:wfc-particle-types}) and tested combinations. Figure columns are different temperatures, curve colors are different particle combinations as shown in {\bf e}. {\bf a)} 3x3x3 wireframe cube yield curves analyzed with a cutoff point of 100\%. {\bf b)} 3x3x3 wireframe cube yield curves analyzed with a cutoff point of 75\%. {\bf c)} 3x3x3 wireframe cube yield curves analyzed with a cutoff point of 50\%. {\bf d)} 3x3x3 wireframe cube yield curves analyzed with a cutoff point of 0\%. Note the falloff of the curves at T=0.01 and 0.04, as partial structures are included in the yield curve but malformed structures are not. {\bf e)} Particle types included in the different simulation groups (curve line colors). Cube colors correspond to schematics in \ref{fig:wfc-particle-types}.} 
       \label{fig:wireframe-cubes}
   \end{figure*}
   
   \begin{figure*}
       \centering
       \includegraphics[width=\textwidth]{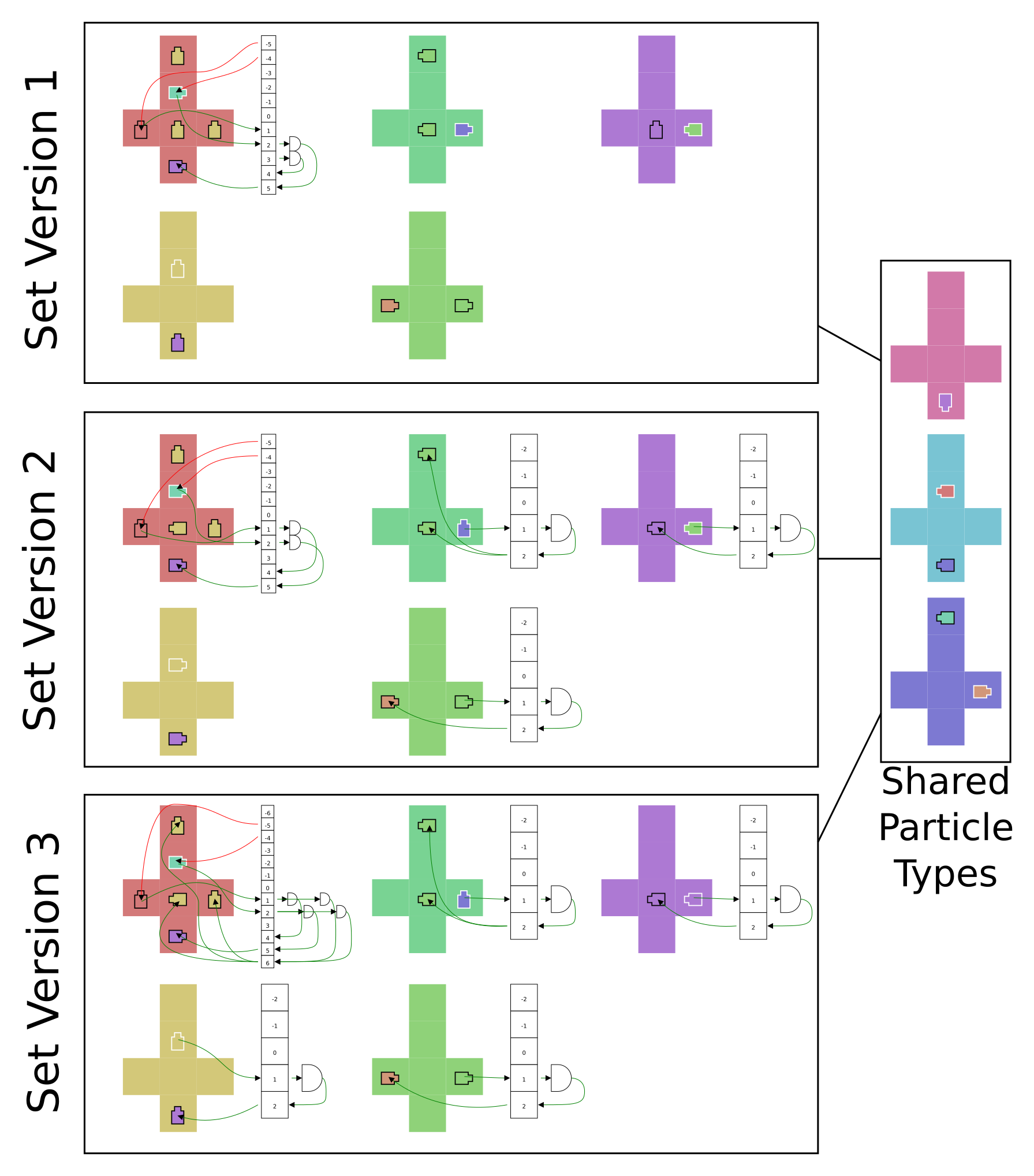}
       \caption{Sets for the wereflamingo. The versions are ordered by increasing complexity. The three sets have the same number of particle types and colors.Version 3 corresponds to the version used in the main text simulations (Fig.~8).}
       \label{fig:wereflamingo-particles-suppl}
   \end{figure*}
   
   \begin{figure*}
       \centering
       \includegraphics[width=\textwidth]{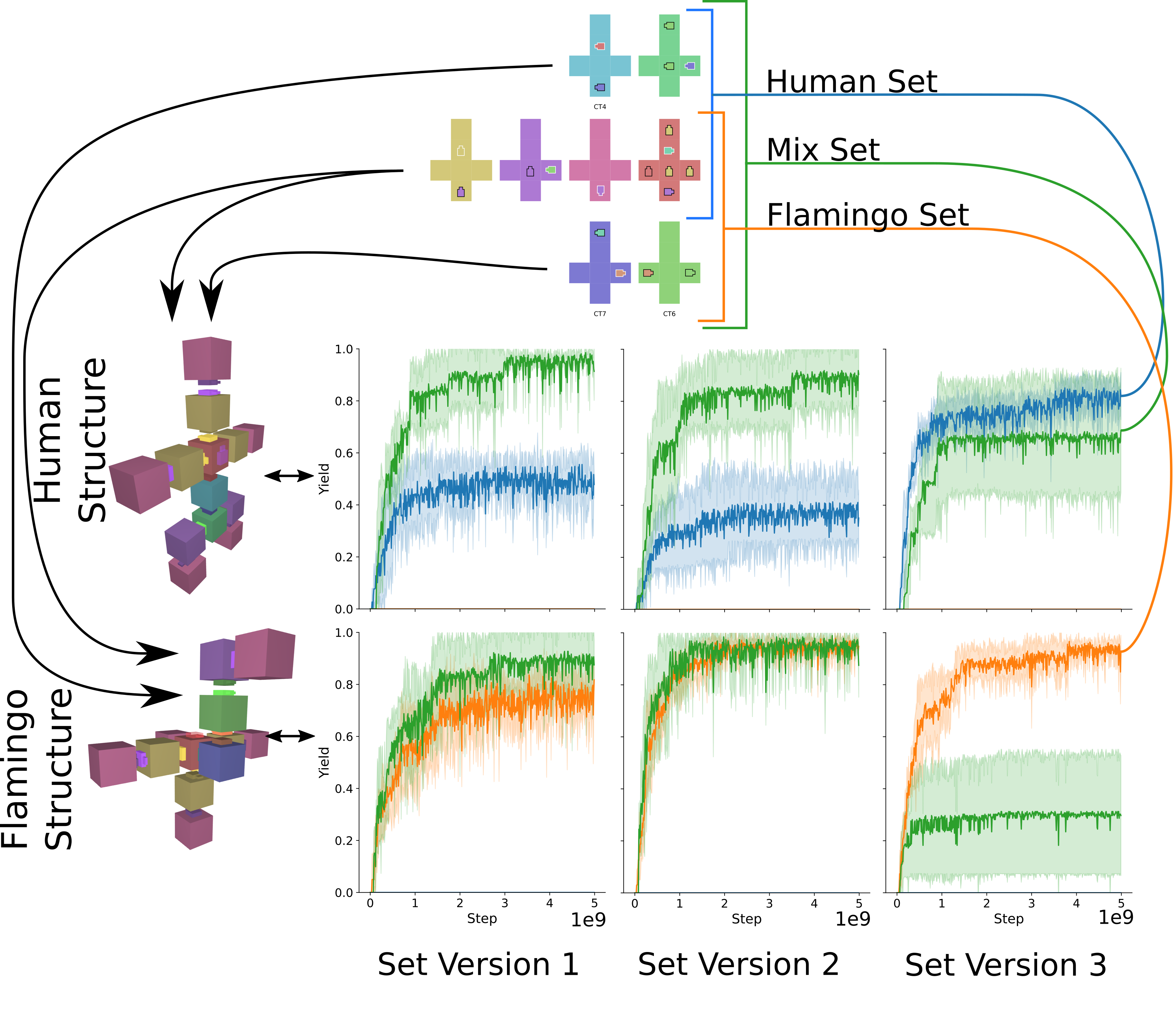}
       \caption{Narrow type 0 wereflamingo yield curves for our three versions (See Fig.~\ref{fig:wereflamingo-particles-suppl}). The color of the lines in the yield curves correspond to the sets of particles included in the yield curves (top, allosteric logic not shown here). The yields were calculated taking into account particle type in order to avoid problems caused by topological similarities between the structures.}
       \label{fig:wereflamingo-suppl}
   \end{figure*}
   
   \begin{figure*}
       \centering
       \includegraphics[width=0.5\textwidth]{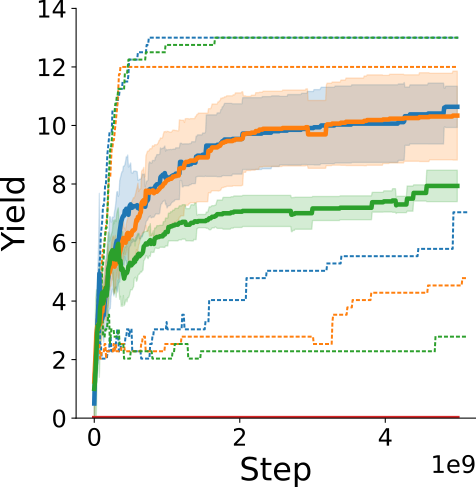}
       \caption{Total cluster sizes in the wereflamingo, version 3 (same base data as Fig.~8). The orange, blue, green, and red colors represent flamingo, human, mix, and neither-assembly particle subsets, as shown in Fig.1~8. The thick lines indicate the mean of the mean cluster sizes of each simulation, with the shaded areas showing the standard deviations. The dotted lines indicate the minima and maxima of the mean cluster sizes of each simulation. The lack of any cluster formations on the red line shows that our "neither-assembly" particle set successfully avoids any cluster formation.}
       \label{fig:wereflamingo-cluster-sizes}
   \end{figure*}
   
   \begin{figure*}
       \centering
       \includegraphics[width=\textwidth]{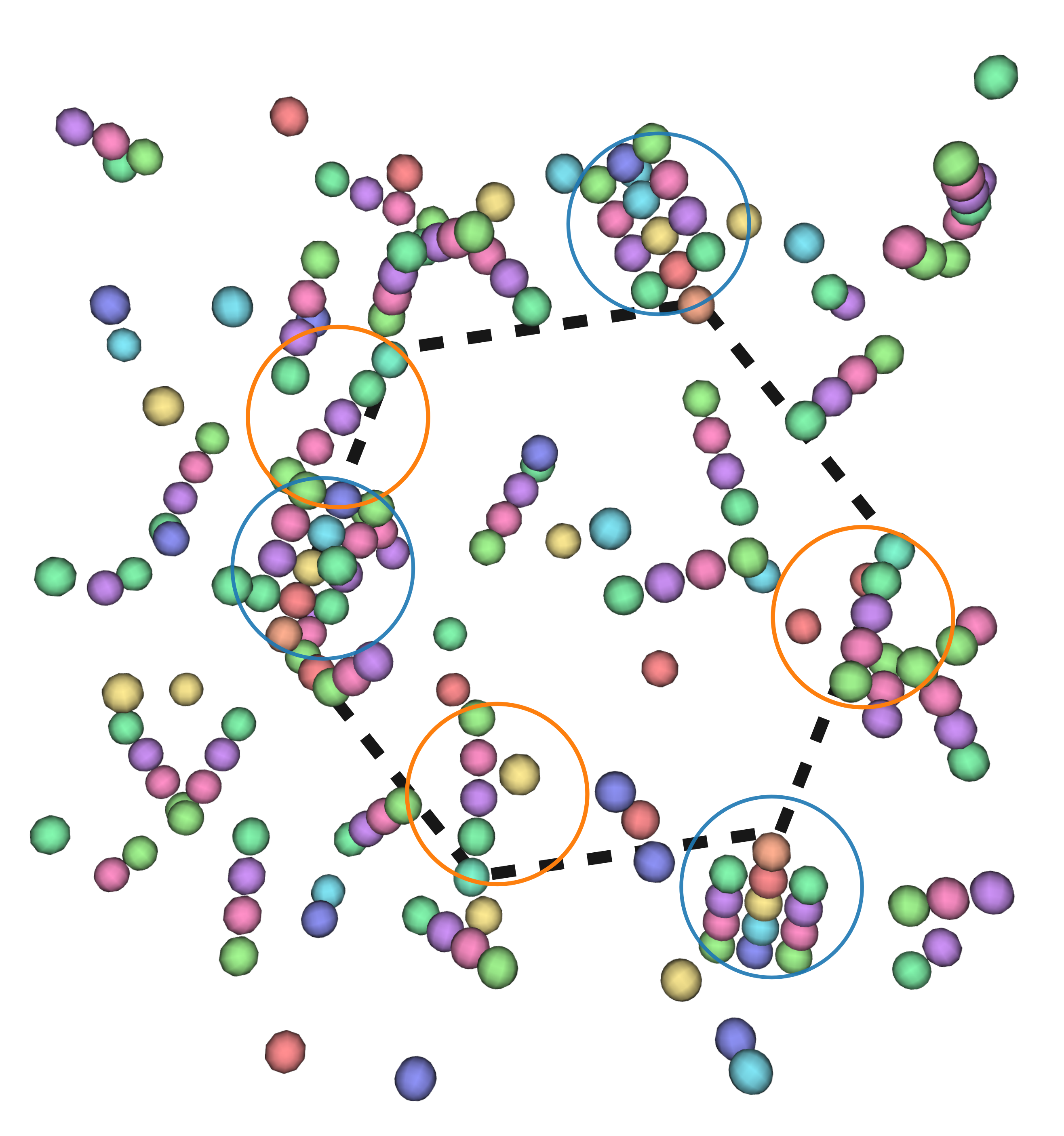}
       \caption{The stonehenge simulation from the main text pre-"wash"}
       \label{fig:stonehenge-blobby-nowash}
   \end{figure*}
   
   \begin{figure*}
       \centering
       \includegraphics[width=\textwidth]{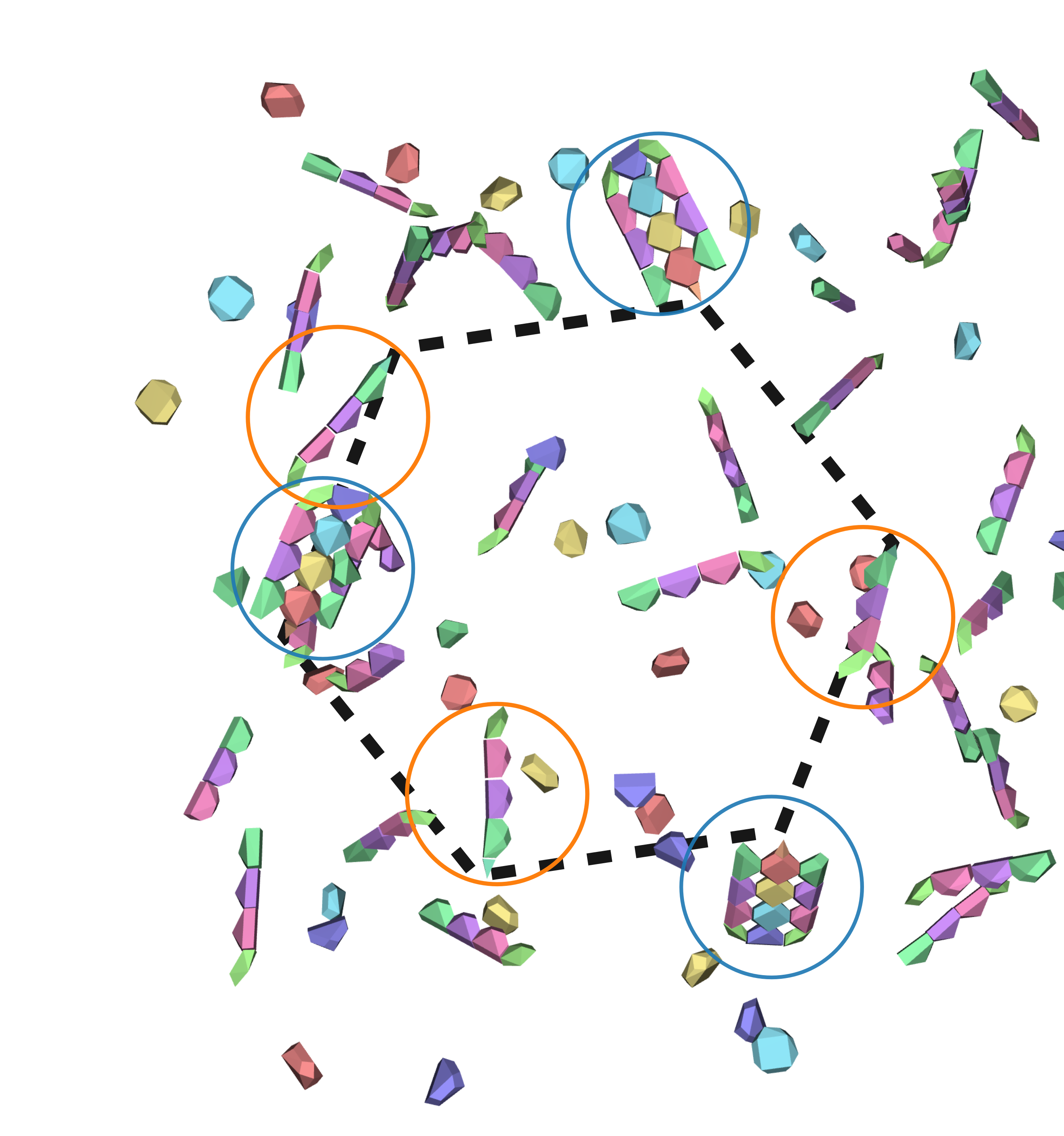}
       \caption{The stonehenge simulation from the main text pre-"wash", visualized with patch-dependant topology. The particles were modeled as spheres but for visualization purposes each patch is treated as a polygon, so each particle is a polyhedron.}
       \label{fig:stonehenge-pointy-nowash}
   \end{figure*}
   
   
\clearpage

\bibliographystyle{aip}
\bibliography{refmain,josh-refs,additional-refs}

\end{document}